\documentclass[
  twocolumn,
  prb,
  showpacs,
  amsmath,
  amssymb,
  superscriptaddress,
  floatfix
]{revtex4}

\usepackage{bm}
\usepackage{dsfont}
\usepackage{graphicx}
\usepackage{pifont}
\usepackage{textcomp}
\usepackage{gensymb}
\usepackage{accents}
\usepackage{upgreek}

\makeatletter
\def\NAT@bibsetnum#1{%
 \setlength{\topsep}{\z@}%
 \NATx@bibsetnum{#1}%
}%
\renewenvironment{thebibliography}[1]{%
 \NAT@thebibliography{#1}%
 \@clubpenalty\clubpenalty
 \let\@TBN@opr\present@bibnote
 \@FMN@list
}{%
 \@endnotesinbib
 \edef\@currentlabel{\arabic{NAT@ctr}}%
 \NAT@endthebibliography
 \global\let\auto@bib\@empty   
}
\newcommand*{\supplementarystart}{%
  \close@column@grid%
  \clearpage%
  \onecolumngrid%
  \setcounter{enumiv}{0} % resets counter for references
  \setcounter{equation}{0} % resets counter for equations
  \setcounter{figure}{0} % resets counter for figs
  \setcounter{table}{0} % resets counter for tables
  \setcounter{page}{1}
  \c@secnumdepth=4
  \renewcommand{\theequation}{s\arabic{equation}} % equations numbered with S...
  \renewcommand{\bibnumfmt}[1]{[s##1]} % bibtems [S...]
  \renewcommand{\@onlinecite}{s\citealp} % citations [S...]
  \renewcommand{\cite}[1]{{[}\onlinecite{##1}{]}}
  \renewcommand{\thefigure}{s\arabic{figure}}
  \renewcommand{\thetable}{s\Roman{table}}
  \renewcommand{\thepage}{s\arabic{page}}
}
\makeatother

\renewcommand{\cite}[1]{{[}\onlinecite{#1}{]}}

\newcommand{\s}{\sum\limits}

\newcommand{\pa}{\partial}

\newcommand{\be}{\begin{equation}}
\newcommand{\e}{\end{equation}}
\newcommand{\beml}{\begin{subequations}}
\newcommand{\eml}{\end{subequations}}
\newcommand{\beq}{\begin{eqnarray}}
\newcommand{\eq}{\end{eqnarray}}
\newcommand{\ba}{\begin{array}}
\newcommand{\ea}{\end{array}}
\newcommand{\bpm}{\begin{pmatrix}}
\newcommand{\epm}{\end{pmatrix}}
\newcommand{\bc}{\begin{cases}}
\newcommand{\ec}{\end{cases}}
\newcommand{\lt}{\left}
\newcommand{\rt}{\right}
\newcommand{\n}{\nonumber}
\newcommand{\la}{\langle}
\newcommand{\ra}{\rangle}
\newcommand{\ep}{\varepsilon}

\newcommand{\bb}{\boldsymbol}
\newcommand{\h}{^\dagger}
\newcommand{\0}{^\phantom{\dagger}}

\DeclareMathOperator{\tr}{Tr}

\begin{document}

\title{Microscopic theory of spin-orbit torques and skyrmion dynamics}

\author{I.\,A.~Ado}
\affiliation{Radboud University, Institute for Molecules and Materials, NL-6525 AJ Nijmegen, The Netherlands}
\author{Oleg~A. Tretiakov}
\affiliation{Institute for Materials Research, Tohoku University, Sendai 980-8577, Japan}
\affiliation{School of Natural Sciences, Far Eastern Federal University, Vladivostok 690950, Russia}
\author{M.~Titov}
\affiliation{Radboud University, Institute for Molecules and Materials, NL-6525 AJ Nijmegen, The Netherlands}

\begin{abstract}
We formulate a general microscopic approach to spin-orbit torques in thin ferromagnet/heavy-metal bilayers in linear response to electric current or electric field. The microscopic theory we develop avoids the notion of spin currents and spin-Hall effect. Instead, the torques are directly related to a local spin polarization of conduction electrons, which is computed from generalized Kubo-St\v{r}eda formulas. A symmetry analysis provides a one-to-one correspondence between polarization susceptibility tensor components and different torque terms in the Landau-Lifshitz-Gilbert equation for magnetization dynamics. The spin-orbit torques arising from Rashba or Dresselhaus type of spin-orbit interaction are shown to have different symmetries. We analyze these spin-orbit torques microscopically for a generic electron model in the presence of an arbitrary smooth magnetic texture. For a model with spin-independent disorder we find a major cancelation of the torques. In this case the only remaining torque corresponds to the magnetization-independent Edelstein effect. Furthermore, our results are applied to analyze the dynamics of a Skyrmion under the action of electric current. 
\end{abstract}

\maketitle

\section{Introduction}

Electrons in a thin layer of a heavy metal (HM) are subject to a large spin-orbit interaction, which couples electron orbital and spin degrees of freedom \cite{
Rashba-SOI,Hoffmann13,review-Rashba1,review-Rashba2,Sinova15}. In a ferromagnet/heavy-metal bilayer the electron spin is also coupled locally to the magnetic moment in the ferromagnet (FM) by means of exchange interaction. Simultaneous presence of these two interactions provides a way to manipulate spin textures in a ferromagnet by means of spin-orbit torques \cite{Garello-switching,SOT1,SOT2,SOT3,SOT4,anti-damping SOT1,anti-damping SOT2,anti-damping SOT-QKE,SOT AFM1,SOT_Manchon2015,SOT AFM2,Emori13,Ryu13,SOT exp,SOT DW Miron1,SOT DW Miron2,SOT exp Klaui,SOT exp Miron1,SOT exp Miron2,SOT exp Miron3}. 

Spin-orbit torques have been indeed recognized recently as a very efficient way to drive ferromagnetic domains with an electric current \cite{Garello-switching,SOT1,SOT2,SOT3,SOT4,anti-damping SOT1,anti-damping SOT2,anti-damping SOT-QKE,SOT AFM1,SOT_Manchon2015,SOT AFM2,Emori13,Ryu13,SOT exp,SOT DW Miron1,SOT DW Miron2,SOT exp Klaui,SOT exp Miron1,SOT exp Miron2,SOT exp Miron3,Liu12,co-pt,Tomasello2014, Linder13,SOTreview1,SOTreview2,SOTreview3}. The effect has been demonstrated recently in ferromagnet/heavy-metal bilayer Ta/CoFeB \cite{Hoffmann15,Hoffmann16} as well as in Pt/Co/Ta and Pt/CoFeB/MgO multilayers \cite{Beach2016} for magnetic Skyrmions. Despite its importance for creating novel magnetic memory devices \cite{Jonietz10,Bader15} the theoretical understanding of current-induced magnetic texture dynamics due to spin-orbit torques remains, however, largely phenomenological. 

In this paper we introduce a systematic approach to spin-orbit torques which can be applied for microscopic analysis of spin-texture dynamics in ferromagnet/heavy-metal bilayers and in a more broad context. Namely, our methodology is straightforward to apply for the computation of both spin-orbit and spin-transfer torques, electron contributions to Gilbert damping, and related quantities in both ferromagnet/HM and antiferromagnet/HM bilayers. It is interesting to note that the microscopic theory developed in this paper completely avoids the notions of spin current and spin-Hall effect \cite{SHEsinova}. 

%%%%%%%%%%%%%%%%%%%%%%%%%%%%
%%%% fig:fig1
%%%%%%%%%%%%%%%%%%%%%%%%%%%%
\begin{figure}[bt]
\includegraphics[width=\columnwidth]{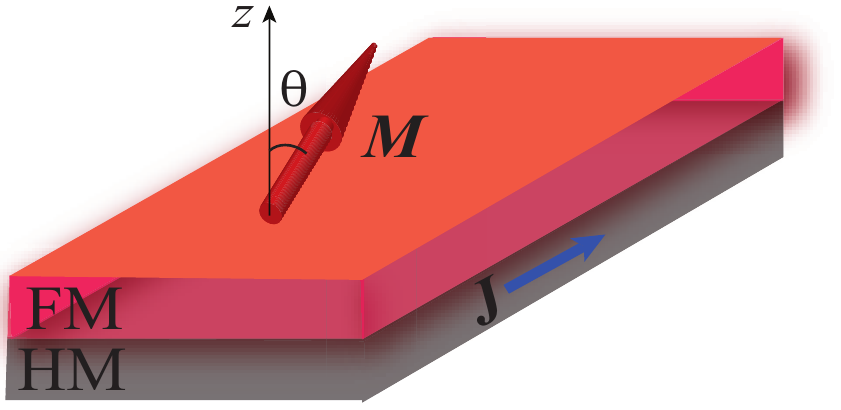}
\caption{Schematic of the model setup. Spin-orbit torque induces dynamics of the magnetization $\mathbf{M}$ in a ferromagnet (FM)/heavy-metal (HM) bilayer under the applied current $J$.}
\label{fig:setup}
\end{figure}
%%%%%%%%%%%%%%%%%%%%%%%%%%%%

In this work we employ a self-consistent mean field approach to the treatment of magnetization dynamics in a ferromagnet/HM bilayer, which is schematically depicted in Fig.~\ref{fig:setup}. We assume that the magnetization dynamics can be described by a classical field $\mathbf{m}(\bb{r},t)$ with the constraint $|\bb{m}|=1$. The unit vector field $\bb{m}$ points in the direction of the locally averaged magnetic moment. (In such a continuous model one does not distinguish individual atomic moments on a lattice.) In this continuous approach the magnetic subsystem, consisting of localized magnetic moments of the ferromagnet, can be described by a classical free energy functional $F[\bb{m}(\bb{r},t)]$, which takes into account all possible magnetic interactions (such as magnetic exchange, anisotropy terms, and Dzyaloshinskii-Moriya interactions) in the ferromagnet, but ignores the effects of the conduction electrons. The latter are described on the basis of an $s$-$d$-like model, which takes into account the exchange coupling between classical magnetic moments (e.g. given by localized $d$-electrons) and the spins of conduction electrons (e.g. $s$-electrons) by means of the following term in the Hamiltonian 
\be
H_{ex}= - J_{\rm{ex}}\, \bb{m} \cdot \bb{\sigma},
\e
where $J_{\rm{ex}}$ is the corresponding exchange energy constant and $\bb{\sigma}=(\sigma_x,\sigma_y,\sigma_z)$ is the vector of Pauli matrices representing spin operators of conduction electrons. The conduction electrons in the FM/HM bilayer are described by an effective Hamiltonian of the form
\be
\label{themodel}
H=\xi_{\bb{p}}+H_{so}+H_{ex}+V(\bb{r}),
\e
where $\bb{p}$ is the momentum operator of electrons, the operator $\xi_{\bb{p}}$ corresponds to the kinetic energy of electrons (in the simplest model $\xi_{\bb{p}}=p^2/2m_e$, where $m_e$ is the effective electron mass), the term $H_{so}$ is responsible for the spin-orbit interaction, and $V(\bb{r})$ represents a disorder potential for conduction electrons. We assume the conducting layer of HM to be thin compared to the electron mean free path, so that the motion of electrons can be considered two-dimensional (2D) in the plane perpendicular to  the interface. Another justification for considering 2D electron transport is that even in somewhat thicker HM layers only the electrons close to magnetization $\bb{m}$ in the FM (interfacial layer close to the FM) can contribute to spin-orbit torques. In the following we consider the spin-orbit interaction of two different types: (i) 2D Rashba spin-orbit interaction corresponding to $H_{so}= \alpha_\textrm{so} (\bb{\sigma}\times\bb{p})_z= \alpha_\textrm{so} (\sigma_xp_y-\sigma_yp_x)$ and (ii) 2D Dresselhaus spin-orbit interaction which corresponds to $H_{so}= \alpha_\textrm{so} (\sigma_x p_x-\sigma_y p_y)$ in a given reference frame. Note that the Rashba type of spin-orbit interaction singles out the direction of the vector $\hat{\bb{z}}$ perpendicular to the plane, while the Dresselhaus type of spin-orbit interaction is defined with respect to the lattice orientation in the $x$-$y$ plane. 

In this paper we will be concerned with the magnetization dynamics, i.e. the dynamics of the vector field $\bb{m}(\bb{r},t)$. Due to the constraint $|\bb{m}|=1$, such a dynamics always yields the equation of the form
\be
\label{first}
\frac{\pa \bb{m}}{\pa t} = \bb{f}\times \bb{m},
\e
where vector $\bb{f}$ has, in general, a functional dependence on $\bb{m}(\bb{r},t)$ and on external fields. In our model we will naturally distinguish two major contributions to vector $\bb{f}$: one originating in classical magnetic moments ($d$-electrons), which are localized in a FM layer, and the other originating in conduction $s$-electrons, which are mainly concentrated in the heavy-metal layer. Hence we can write
\be
\label{ffield}
\bb{f}(\bb{r},t)=\gamma \bb{H}_{\textrm{eff}} +\kappa \bb{s},
\e
where $H_{\textrm{eff}}(\bb{r},t)= -\delta F[\bb{\mathcal{M}}]/\delta \bb{\mathcal{M}}(\bb{r},t)$ is the effective magnetic field created by the localized moments in the ferromagnet ($\bb{\mathcal{M}} =  |\bb{\mathcal{M}}|\bb{m}$ is the magnetization), whereas vector $\bb{s}(\bb{r},t)$ is the non-equilibrium electron polarization density induced by conduction electrons. Here we introduce the gyromagnetic ratio $\gamma$ for the spins in the ferromagnet and coefficient $\kappa=(g\mu_B)^2\mu_0/d$ defined by the electron $g$-factor ($g=2$), Bohr magneton $\mu_B$, vacuum permeability $\mu_0$, and the effective thickness of conduction layer $d$. Throughout the paper we set the Planck constant and the speed of light to be unity, $\hbar=c=1$.

In the mean field approach we consider conduction electrons in the presence of both non-equilibrium classical field $\bb{m}(\bb{r},t)$ and electric field $\bb{E}(t)$ to obtain the corresponding non-equilibrium spin polarization density $\bb{s}(\bb{r},t)$. The relation between $\bb{s}$, magnetization $\bb{m}$, and electric field $\bb{E}$ is generally non-local both in time and in space on the scales of the electron scattering time $\tau$ and electron mean free path $\ell$, respectively. Assuming that $\bb{m}$ and $\bb{E}$ are slow and smooth on these electronic scales, one may justify the gradient expansion that takes into account the non-locality in an approximate manner. In this case one can expand $\bb{s}$ as follows:
\be
\label{sEq}
s_\alpha= K_{\alpha\beta} E_\beta + R^{\gamma\delta}_{\alpha\beta} E_{\beta}\nabla_\delta m_{\gamma} + u\,\pa_t m_\alpha+\dots,
\e
where the summation over repeated indices is assumed. According to the widely accepted classification \cite{STTreview1,STTreview2,STTreview3}, one should associate tensor $\hat{K}$ with the so-called spin-orbit torques (SOT) and tensor $\hat{R}$ with the so-called spin-transfer torques (STT) for the in-plane current geometry.  Furthermore, the coefficient $u$ in Eq.~(\ref{sEq}) defines the conduction-electron contribution to the Gilbert damping. Clearly the decomposition of Eq.~(\ref{sEq}) may be further detailed by considering terms containing, e.\,g. both the time derivative of $\bb{m}$ and electric field. Such a term would correspond to a ``torque'' on the Gilbert damping. Below we will focus specifically only on the analysis of the first term in Eq.~(\ref{sEq}), i.e. on tensor $\hat{K}$ defining SOTs. A simple symmetry argument suggests that the SOTs are vanishing in the absence of spin-orbit interaction.

In order to compute SOT microscopically we restrict ourselves to the calculation of non-equilibrium spin-polarization that does not involve any gradients of magnetization, $\bb{s}= \hat{K}\bb{E}$, and define the corresponding SOT as $\bb{T}=\kappa\, \bb{s}\times\bb{m}$. Note that in the absence of spin-orbit interaction, the spin-polarization density $\bb{s}$ is macroscopically large (proportional to the number of electrons) in the direction of magnetization $\bb{m}$. This may be seen as a diffusion pole in the corresponding diagrammatic calculation below that calls for an accurate analysis of the so-called vertex corrections. To the best of our knowledge, this technical difficulty has never been accurately considered even for the simplest models. 

The plan of the paper as follows. In Sec.~\ref{sec:Rashba} we present a general symmetry analysis of torques for a 2D $s$-$d$ model with Rashba spin-orbit interaction, which is sometimes referred to as Bychkov-Rashba model \cite{Bychkov84}. It is shown that the SOTs are directly related to a susceptibility tensor $K$ that defines local non-equilibrium polarization of conduction electrons. Then the symmetry analysis is supplemented by the microscopic calculation of spin-orbit torques for a particular case of quadratic dispersion and Gaussian white-noise disorder, which is taken into account in the self-consistent Born approximation. For this particular model we prove the full cancelation of three out of four torques, while the remaining torque is shown to be reduced to the magnetization-independent Edelstein effect. Even though the exact cancelation is absent in more complex models, our analysis suggests that a strong suppression of spin-orbit torques that are non-linear in magnetization $\bb{m}$ is generic in two dimensions. In Sec.~\ref{sec:Dressel} we consider a model with the spin-orbit interaction of Dresselhaus type. We demonstrate that the torques have completely different symmetries in this case, but the torque coefficients in this model can be directly related to those already defined for the Rashba model. Hence no separate calculation is necessary to fully describe spin-orbit torques in the Dreselhaus model. In Sec.~\ref{sec:skyrm} we consider the motion of Skyrmions under the action of a small electric current in both Rashba and Dresselhaus model in the presence of all possible spin-orbit torques. This analysis is based on the generalized Thiele equation. We summarize our results in Sec.~\ref{sec:conclusion}.  

\section{SOT in Rashba model}
\label{sec:Rashba}

\subsection{Symmetry analysis}

The Landau-Lifshitz-Gilbert (LLG) equation \cite{STTreview1,STTreview2,STTreview3} follows directly from Eqs.~(\ref{first},\ref{ffield}) in the form
\be
\label{LLGmain}
\frac{\pa \bb{m}}{\pa t} = -\gamma\, \bb{m}\times \bb{H}_{\textrm{eff}} +\alpha_\textrm{G}\, \bb{m}\times \frac{\pa \bb{m}}{\pa t}+\bb{T}, 
\e
where $\alpha_G=\kappa\,u$ is the electron-induced Gilbert damping constant, derivation of which falls out of the scope of the present paper (in general, $\alpha_G$ is a phenomenological constant in this equation, which has contributions as well from other mechanisms, such as from phonons etc.), and $\bb{T}= \kappa\, \bb{s}\times\bb{m}$ with $\bb{s}=\hat{K} \bb{E}$ is the non-equilibrium spin-polarization due to the electric field. The latter is related to the electric current by means of the inverse conductivity tensor.

Even before any microscopic calculation is performed, a straightforward symmetry analysis can be applied to reconstruct the symmetries of possible spin-orbit torques arising in this model. In the particular case of Rashba spin-orbit interaction we arrive at the following expression for the electric-field driven spin-orbit torques $\bb{T}=\bb{T}^\perp+\bb{T}^\parallel$ \cite{SOT exp Miron3},
\beml
\label{torques}
\begin{align}
\bb{T}^\parallel &= a\,\bb{m}\times (\hat{\bb{z}}\times \bb{E}) + c\, \bb{m} \times (\bb{m}\times \hat{\bb{z}})\,(\bb{m}\cdot \bb{E}), \\
\bb{T}^\perp &= b\,\bb{m}\times \lt(\bb{m}\times (\hat{\bb{z}}\times \bb{E})\rt) + d\, \bb{m} \times\hat{\bb{z}}\,(\bb{m}\cdot \bb{E}),
\end{align}
\eml
where $\bb{E}$ is the in-plane electric field, $\hat{\bb{z}}$ is the unit vector in $z$ direction (which is perpendicular to the 2D plane of electron gas, see Fig.~\ref{fig:setup}), and the quantities $a$, $b$, $c$, and $d$ are analytic functions of $(\bb{m}\cdot \hat{\bb{z}})^2$, i.\,e. the functions of $\cos^2\theta$, where $\theta(\bb{r},t)$ is the local angle between vectors $\hat{\bb{z}}$ and $\bb{m}(\bb{r},t)$. Since electric field is invariant under the time reversal we have to regard $\bb{T}^\parallel$ as the damping-like (dissipative) torque, which changes sign under the time reversal, whereas $\bb{T}^\perp$ has to be regarded as the field-like (dissipationless) torque, which is invariant under the time reversal. 

Since vector $\bb{T}$ is perpendicular to $\bb{m}$ by construction, it may always be decomposed using two non-collinear vectors in the plane perpendicular to $\bb{m}$. Thus, the result of Eqs.~(\ref{torques}) can always be rewritten in the form $\bb{T}= \tilde{a}\,\bb{m}\times (\hat{\bb{z}}\times \bb{E}) +\tilde{b}\,\bb{m}\times \lt(\bb{m}\times (\hat{\bb{z}}\times \bb{E})\rt)$. However, the disadvantage of this representation is in the complex dependence of coefficients $\tilde{a}$ and $\tilde{b}$ on magnetization $\bb{m}$, which makes them neither even nor odd functions of time. Thus, we find the representation of Eq.~(\ref{torques}) more natural for 2D Rashba model because all coefficients $a$, $b$, $c$, and $d$ can also be shown to become constants in the good metal limit (i.\,e. in the limit $\ep_F\tau \gg 1$, where $\ep_F$ is the Fermi energy), irrespective of the model chosen for the disorder. 

To justify Eqs.~(\ref{torques}) we perform a symmetry analysis of the Bychkov-Rashba model \cite{Bychkov84}. For convenience, we fix the reference frame such that the $x$ direction is chosen by the projection of the vector $\bb{m}$ on the 2D plane, hence the model of Eq.~(\ref{themodel}) reads
\be
\label{Rashba}
H=\xi_{\bb{p}}+\alpha_\textrm{so} \, (\bb{\sigma} \times \bb{p})_z - M_x \sigma_{x} - M_z \sigma_z+V(\bb{r}),
\e
where $\xi_{\bb{p}}$ is an isotropic electron dispersion, $\bb{M}=J_\textrm{ex}\,\bb{m}$ is the vector of exchange field, and $\sigma_\alpha$ are the Pauli matrices that represent electron spin operators. The model (\ref{Rashba}) includes all the key ingredients: the spin-orbit coupling of the strength $\alpha_\textrm{so}$ and the exchange coupling between conduction electron spins in the heavy metal and localized moments of the ferromagnet. Note that the in-plane component of the exchange field $\bb{M}$ introduces an anisotropy for 2D electrons that is fully taken into account in our subsequent analysis. 

We assume that magnetic texture is smooth on the scale of the electron mean free path, hence a gradient expansion with respect to $\nabla_\alpha \bb{m}$ is justified. As we already noted we focus below on the spin-orbit torques, which appears in zero (leading) order of the gradient expansion. The corresponding non-equilibrium spin density that is formed in the bulk of the sample in a response to the electric field is given by  $\bb{s} = \hat{K} \bb{E}$, where $\hat{K}$ is a 6-component susceptibility tensor. This tensor is defined (up to a topological contribution discussed in the Appendix \ref{app:Kubo}) by the generalized Kubo-St\v{r}eda formula \cite{Streda},
\begin{align}
\label{K}
K_{\alpha \beta}=&\frac{e}{8\pi} \int\frac{d^2\bb{p}}{(2\pi)^2}\tr\big[\sigma_\alpha \lt(G^R_{\bb{p}} -G^A_{\bb{p}}\rt)v_{\beta} G^A_{\bb{p}}\n\\ 
&- \sigma_\alpha G^R_{\bb{p}} v_{\beta} \lt(G^R_{\bb{p}} -G^A_{\bb{p}}\rt)\big],
\end{align}
where $e$ is the electron charge, $\bb{v}=\nabla_{\bb{p}}\xi_p+\alpha_\textrm{R}  \hat{\bb{z}}\times \bb{\sigma}$ is the in-plane electron velocity operator and $G^R_{\bb{p}}=[\ep-H_{\bb{p}}+i0]^{-1}$ is the retarded Green's function with the Fermi energy $\ep$ (for a sake of the symmetry analysis we avoid first the detailed consideration of the disorder and formally set $V=0$; see Appendix \ref{app:Kubo} for more details). 

Symmetry properties of $\hat{K}$ can be readily established from Eq.~(\ref{Rashba}) with the help of the following symmetry transformations
\be
\label{sym}
\sigma_x H[-p_x] \sigma_x = H[-m_z],\;\; \sigma_z H[-\bb{p}] \sigma_z = H[-m_x],
\e
where the notation $H[-p_x]$ stands, for example, for the Hamiltonian $H$ with the substitution $p_x\to -p_x$.  Applying the same transformations to the velocity operator we find the relations $\sigma_x v_x[-p_x] \sigma_x = -v_x$,  $\sigma_x v_y[-p_x] \sigma_x = v_y$, and $\sigma_z \bb{v}[-\bb{p}] \sigma_z = -\bb{v}$. We now undertake the change of variables $p_x\to -p_x$ or $\bb{p}\to -\bb{p}$ under the integral in Eq.~(\ref{K}). Then, we apply the corresponding symmetry transformations of Eq.~(\ref{sym}) to the Green's functions and to the velocity and spin operators. In this way we figure out if a given $\hat{K}$-tensor component is even or odd function of $m_x$ and $m_z$. The resulting symmetry relations can be expressed as
\be
\label{Ksym}
\hat{K}= \frac{1}{\kappa}\bpm m_z \kappa_{xx} & \kappa_{xy}\\  \kappa_{yx} & m_z \kappa_{yy} \\  m_x \kappa_{zx} & m_x m_z \kappa_{zy}\epm,
\e
where $\kappa_{\alpha\beta}$ are some analytic functions of $m_z^2=1-m_x^2$, i.\,e.\,they do not change sign under the transformations $m_x\to -m_x$ or $m_z\to -m_z$. Using that $\bb{T}=\kappa\, \bb{s}\times\bb{m}$, we confirm the ansatz of Eq.~(\ref{torques}) and establish the following relations,
\beml
\label{symmetries}
\begin{align}
&a=\kappa_{xy}-m_x^2 \kappa_{zy},\quad c=(\kappa_{xy}+\kappa_{yx})/m_x^2- \kappa_{zy},\\
&b=-\kappa_{yy},\quad\;\; d=\kappa_{xx}+\kappa_{zx}+(\kappa_{yy}-\kappa_{xx})/m_x^2,
\end{align}
\eml
which connect spin-orbit torques of Eqs.~(\ref{torques}) with electron spin susceptibilities $K_{\alpha\beta}$ defined in the special reference frame of Eq.~(\ref{Rashba}). 

In an experiment, it is not the electric field $\bb{E}$ which is applied to FM/HM bilayer but rather the electric current $\bb{J}=\hat{\sigma}\bb{E}$,  where $\hat{\sigma}$ stands for the conductivity tensor. The latter is also defined by the standard Kubo formula, that is analogous to Eq.~(\ref{K}),
\begin{align}
\label{sigma}
\sigma_{\alpha \beta}=\;&\frac{e^2}{4\pi} \int\frac{d^2\bb{p}}{(2\pi)^2}\tr\big[v_\alpha \lt(G^R_{\bb{p}} -G^A_{\bb{p}}\rt)v_{\beta} G^A_{\bb{p}}\n\\ 
&- v_\alpha G^R_{\bb{p}} v_{\beta} \lt(G^R_{\bb{p}} -G^A_{\bb{p}}\rt)\big].
\end{align}
From symmetry transformations of Eq.~(\ref{sym}) we immediately confirm the well-known symmetry properties of the conductivity tensor: the Hall conductivity $\sigma_{xy}$ is an odd function of $m_z$ but even function of $m_x$, whereas the longitudinal conductivity $\sigma_{xx}$ is an even function of both. 

In the case of current-driven magnetization dynamics the resulting spin density, $\bb{s}=\hat{K}_\textrm{J} \bb{J}$, is determined by the tensor $\hat{K}_\textrm{J}=\hat{K}\hat{\sigma}^{-1}$ instead of $\hat{K}$. Similarly to Eq.~(\ref{torques}) the symmetries of $\hat{K}_\textrm{J}$ justify the current-induced torques 
\beml
\label{torques1}
\begin{align}
\!\!\bb{T}_\textrm{J}^\parallel &= a_\textrm{J}\,\bb{m}\times (\hat{\bb{z}}\times \bb{J})  + c_\textrm{J} \bb{m} \times (\bb{m}\times \hat{\bb{z}})\,(\bb{m}\cdot \bb{J}),\\
\!\!\bb{T}_\textrm{J}^\perp &= b_\textrm{J}\,\bb{m}\times \lt(\bb{m}\times (\hat{\bb{z}}\times \bb{J})\rt) +  d_\textrm{J} \bb{m} \times\hat{\bb{z}}\,(\bb{m}\cdot \bb{J}),
\end{align}
\eml
where $a_\textrm{J}$, $b_\textrm{J}$, $c_\textrm{J}$, and $d_\textrm{J}$ are related to the entities of the tensor $\hat{K}_\textrm{J}$ in the same way as $a$, $c$, $b$, and $d$ are related to $\hat{K}$ in Eqs.~(\ref{Ksym}) and (\ref{symmetries}). For a sake of completeness we write down these relations explicitly in Appendix~\ref{app:KJ}. Since $\bb{J}$ changes sign under the time reversal, the torque classification is now reversed as compared to Eq.~(\ref{torques}), namely, $\bb{T}_\textrm{J}^\parallel$ is even under the time reversal, and hence it has to be regarded as the field-like torque, whereas $\bb{T}_\textrm{J}^\perp$ is odd under the time-reversal, hence it is the damping-like torque. 

%%%%%%%%%%%%%%%%%%%%%%%%%%%%
%%%% fig:fig1
%%%%%%%%%%%%%%%%%%%%%%%%%%%%
\begin{figure}[bt]
\includegraphics[width=\columnwidth]{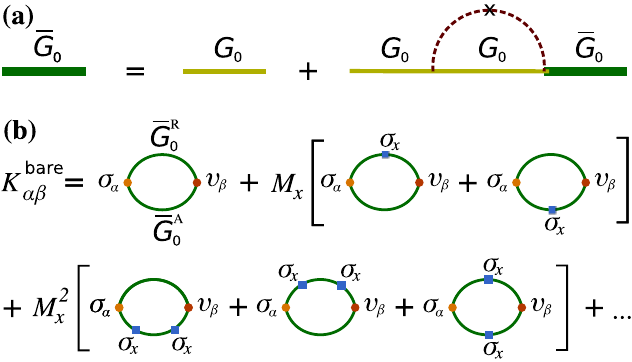}
\caption{(a) Diagrammatic representation of the Born approximation used; (b) Expansion of the ``bare'' spin susceptibility tensor $\hat{K}^\textrm{bare}$ in powers of in-plane magnetization component.}
\label{fig1}
\end{figure}
%%%%%%%%%%%%%%%%%%%%%%%%%%%%%

The results given by Eqs.~(\ref{K}), (\ref{Ksym}), (\ref{symmetries}), and (\ref{torques1}) provide a general microscopic framework to analyze spin-orbit torques in a ferromagnet/heavy-metal bilayer with Rashba spin-orbit interaction.  We stress that our theoretical construction completely avoids the notion of spin current and spin-Hall effect since these concepts appear not to be necessary for the description of spin-orbit torques. Our theory also generalizes previous works on the subject \cite{Bijl12,Hals13,Lee15}.

\subsection{Microscopic analysis}

Let us now compute the SOT microscopically for a widely used Bychkov-Rashba model that is given by Eq.~(\ref{Rashba}) with $\xi_{\bb{p}}=p^2/2m_e$. In order to capture generic behavior of SOT in a heavy metal we consider the case of Gaussian spin-independent disorder, that is characterized by the correlators $\la V(\bb{r}) V(\bb{r}') \ra = (m_e\tau)^{-1}\delta(\bb{r}-\bb{r}')$ and $\la V(\bb{r}) \ra=0$, where brackets stand for disorder averaging and $\tau$ is the scattering time. For potential $V(\bb{r})=V_0\sum_i\delta(\bb{r}-\bb{R}_i)$ with the uniformly distributed impurity coordinates $\bb{R}_i$, one finds the relation $n_{\textrm{imp}}V_0^2 = (m_e\tau)^{-1}$, where $n_{\textrm{imp}}$ is the 2D impurity concentration. 

The limit of Gaussian disorder formally corresponds to the limit $V_0\to 0$ and $n_{\textrm{imp}}\to \infty$, such that the scattering time $\tau$ remains constant.  The limit of a good metal assumes also sufficiently large Fermi energy, $\ep > E^*$ \cite{footnote}, which corresponds to the two spin-split Fermi surfaces.  In this energy band the topological contribution to $\hat{K}$ (given by Eq.~(\ref{KII}) in Appendix \ref{app:Kubo}) and the analogous contribution to the Hall conductivity $\sigma_{xy}$ vanish due to the vanishing Berry curvature.

The difficulty of the microscopic analysis is mostly due to the in-plane anisotropy of the model (\ref{Rashba}) that is caused by the in-plane component of the exchange field $\bb{M}$. We treat this anisotropy perturbatively with the help of the Dyson equation $G=G_0-M_x \,G_0 \sigma_x G$, where $G_0$ refers to the Green's function taken at $M_x=0$. The disorder averaged tensors $\hat{K}$ and $\hat{\sigma}$ are then calculated in each order with respect to $M_x$ within the non-crossing (diffusive) approximation, which is equivalent in this case to the self-consistent Born approximation. Eventually we establish some exact relations that allow for the exact summation of the perturbation series in all orders with respect to the anisotropy. 

We start with the ``bare'' contributions to $\hat{K}$ and $\hat{\sigma}$ shown in Fig.~\ref{fig1}(b). Those are obtained from Eqs.~(\ref{K}) and~(\ref{sigma}) by replacing Green's functions $G^{R,A}$ in Eq.~(\ref{K}) with the corresponding disorder-averaged Green's functions $\bar{G}^{R,A}$ in the Born approximation shown in Fig.~\ref{fig1}(a). For $M_x=0$ one simply finds $\bar{G}_0^{R,A}=[\ep-H-\Sigma_0^{R,A}]^{-1}$, where $\Sigma_0^{R,A}=\mp i/2\tau$ is the self-energy for $\ep>M$ \cite{Ado2}. In the model considered the $\sigma_z$ component of the self-energy vanishes leading to some dramatic simplifications that we describe below.

The perturbative expansion with respect to the anisotropy $M_x\sigma_x$ is actually an expansion in powers of the dimensionless parameter $\mu_x=M_x/\Delta_\textrm{S}$, where $\Delta_\textrm{S}=\sqrt{M_z^2+2\ep m_e\alpha_{\textrm{R}}^2}$ is the spin sub-band splitting that itself depends on $M_z$. The direct calculation of the bare tensors up to the terms of the fourth order in $\mu_x$ yields the following expressions
\begin{widetext}
\beml
\label{bare}
\begin{align}
\hat{K}^{\textrm{bare}}=&\,\frac{m_e \alpha_\textrm{R}e}{4\pi\Delta_\textrm{S}} 
\bpm
-\mu_z(1-\mu_x^2(1-2\mu_z^2)) & \tau\Delta_\textrm{S}(1+\mu_z^2+\mu_x^2\mu_z^2(1-3\mu_z^2)),\\
-\tau\Delta_\textrm{S}(1+\mu_z^2+2\mu_x^2\mu_z^2(1-\mu_z^2)) & \mu_z(1+\mu_x^2(1-2\mu_z^2)),\\
-\mu_x\mu_z^2 (1+\mu_x^2(2-3\mu_z^2))& 2\tau\Delta_\textrm{S} \mu_x\mu_z^3(1+\mu_x^2(2-3\mu_z^2))
\epm+\mathcal{O}(\mu_x^4),\\
\hat{\sigma}^{\textrm{bare}}=&\,\frac{e^2}{2\pi} 
\bpm
2\ep\tau\lt( 1+\frac{\Delta_\textrm{S}^2}{4\ep^2}(1-\mu_z^2)^2(1-\mu_z^2\mu_x^2)\rt) & \frac{\Delta_\textrm{S}}{2\ep}\mu_z (1-\mu_z^2)(1+\mu_x^2(1-2\mu_z^2)) \\
-\frac{\Delta_\textrm{S}}{2\ep}\mu_z (1-\mu_z^2)(1+\mu_x^2(1-2\mu_z^2)) & 2\ep\tau\lt( 1+\frac{\Delta_\textrm{S}^2}{4\ep^2}(1-\mu_z^2)(1-\mu_z^2-\mu_z^2\mu_x^2(1-3\mu_z^2))\rt)
\epm+\mathcal{O}(\mu_x^4),
\end{align}
\eml
\end{widetext}
where we introduced $\mu_\alpha=M_\alpha/\Delta_\textrm{S}$. Even though the results of Eqs.~(\ref{bare}) are incomplete (since they do not take into account vertex corrections and are, therefore, not gauge invariant), one can anyway make several useful observations based on them. First of all, the results of Eqs.~(\ref{bare}) are evidently consistent with the symmetry analysis of Eq.~(\ref{symmetries}).  Moreover, for the case $\Delta_\textrm{S} \tau \gg 1$ (i.\,e. for the sub-band splitting much larger than the disorder broadening), the components $K_{xy}$, $K_{yx}$, and $K_{zy}$ are greater than $K_{xx}$, $K_{yy}$, and $K_{xz}$ components. From Eqs.~(\ref{Ksym}) and~(\ref{symmetries}), one concludes that the coefficients $b$ and $d$ are generally smaller than $a$ and $c$ in the limit of well separated sub-bands,  $\Delta_\textrm{S} \tau \gg 1$. In contrast, the anomalous Hall conductivity, $\sigma_{xy}$, is smaller than $\sigma_{xx}$ by the parameter $\ep \tau$ that is large in any metal. 

One may also see that the dependence of the tensors $\hat{K}$ and $\hat{\sigma}$ on the angle $\theta$ ($m_z=\cos\theta$) between the magnetization and the normal to the plane direction, $\hat{\mathbf{z}}$, is negligible in the limit $J_\textrm{ex}\ll \Delta_S$. For large enough Fermi energy, $\ep$, and sufficiently clean system the latter condition is typically fulfilled, and therefore all coefficients $a$, $b$, $c$ and $d$ are generally constant. 

The components $K_{xy}$, $K_{yx}$, $K_{zy}$, $\sigma_{xx}$, and $\sigma_{yy}$ are proportional to the scattering time $\tau$ which reflects their dissipative character. These quantities are diverging in the clean limit $\tau\to \infty$, the behavior which is well-known for the conductivity from Drude theory. The components $K_{xx}$, $K_{yy}$, $K_{xz}$, $\sigma_{xy}$, and $\sigma_{yx}$ represent dissipationless quantities.  In the clean limit they are $\tau$-independent and equal to intrinsic contributions (see Ref.~\cite{Lee15}). The latter is related to the Berry curvature \cite{Berry} in the clean model, i.\,e. in the limit $V\to 0$ or $\tau \to \infty$.  

It is easy to see that the quantities $a$ and $c$, which define the damping-like torque $\bb{T}^\parallel$, are indeed dissipative (proportional to $\tau$), while the quantities $b$ and $d$, which define the field-like torque $\bb{T}^\perp$, are dissipationless ($\tau$-independent). However, it has to be stressed that it is insufficient to calculate the $\tau$-independent correlators $K_{xx}$, $K_{yy}$, $K_{xz}$, $\sigma_{xy}$, and $\sigma_{yx}$ in the non-crossing approximation as was demonstrated explicitly in Refs.~\cite{Ado1,Ado2}. The consistent analysis of such correlators must take into account the skew-scattering on rare impurity configurations  \cite{Ado2}. The self-consistent Born approximation remains, however, fully consistent for the leading-order components that define the coefficients $a$ and $c$ as well as $a_\textrm{J}$ and $c_\textrm{J}$.

Using that $\hat{K}_\textrm{J}=\hat{K}\hat{\sigma}^{-1}$ we find that the quantities $a_\textrm{J}$ and $c_\textrm{J}$ do not depend on $\tau$, while $b_\textrm{J}$ and $d_\textrm{J}$ are inversely  proportional to $\tau$. This is consistent with $\bb{T}_\textrm{J}^\parallel$ identified as the field-like torque and $\bb{T}_\textrm{J}^\perp$ as the damping-like torque.

All these observations made from the incomplete results of Eq.~(\ref{bare}) are certainly general and remain valid for a generic disorder. It is, however, instructive to complete the calculation by adding all non-crossing impurity lines connecting $G^R$ and $G^A$ in the diagrams of Fig.~\ref{fig1}(b) in all orders with respect to $M_x$. The procedure is reduced to the calculation of vertex corrections and diffusions as discussed in  Appendix \ref{app:disorder}.  Remarkably, this procedure leads to the full cancelation of the entire $\bb{M}$ dependence in both $\hat{K}$ and $\hat{\sigma}$ tensors in all orders of the perturbation theory with respect to the anisotropy. The final result in all orders is extraordinary compact:
\be
\label{dressed}
\hat{K}=\frac{e\alpha_\textrm{so} m_e}{4\pi}  \bpm 0 & 2\tau\\ -2\tau& 0\\ 0 & 0 \epm,\quad
\hat{\sigma}=\frac{e^2}{2\pi}2\tau(\ep+m_e\alpha_\textrm{R}^2) \hat{\openone},
\e
and is manifestly independent of vector $\bb{M}$. The only non-vanishing component of $\hat{K}$ represents the so-called Edelstein spin accumulation \cite{Edelstein effect1,Edelstein effect2,Edelstein effect3} or Edelstein effect that is present even for $M=0$, i.\,e. in the absence of the exchange field. From Eq.~(\ref{dressed}) we also obtain
\be
\label{KJ}
\hat{K}_\textrm{J}= \frac{a_\textrm{J}}{\kappa}\bpm 0 & 1\\ -1& 0\\ 0 & 0 \epm,\quad a_\textrm{J}=\frac{\alpha_\textrm{so} m_e \kappa}{2e(\ep+m_e\alpha_\textrm{so}^2)},
\e
which is independent of the scattering time. The results of Eqs.~(\ref{K}) and~(\ref{dressed}) correspond to $a= \tau \kappa \alpha_\textrm{so} e m_e/2\pi$ and $a_\textrm{J}$ given by Eq.~(\ref{KJ}), while all other torques are vanishing $b=c=d=b_\textrm{J}=c_\textrm{J}=d_\textrm{J}=0$. Thus, the only spin-orbit torque that is finite in the self-consistent Born approximation is induced by the magnetization-independent Edelstein effect \cite{Edelstein effect1,Edelstein effect2,Edelstein effect3}.

The remarkable cancelation of the entire dependence of $\hat{K}$ and $\hat{\sigma}$ on the exchange field $\bb{M}$ can be traced back to the vanishing $\sigma_z$ component of the Born self-energy. Similar cancelation of intrinsic contributions by disorder scattering is well known for the spin-Hall effect \cite{Inoue2004}. We note, that the bare (intrinsic) contributions to spin-orbit torques represented by the components $K_{xx}$ and $K_{yy}$ of Eq.~(\ref{bare}) have also been analyzed numerically in Ref.~\cite{Lee15}.

There exist many different ways to overcome the full cancelation. Taking into account skew scattering on rare impurity configuration will lead to finite, though very small, components $K_{xx}$, $K_{yy}$, and $K_{zx}$ \cite{Ado2,Ado3} even within the present model. Generalization of the model to account for strong impurities or, even better, paramagnetic impurities (which scatter electrons with spins parallel and antiparallel to $\bb{m}$ with notably different cross-sections  \cite{anti-damping SOT-QKE}) also leads to the absence of exact cancellations \cite{Alireza}. Finally, the presence of additional in-plane anisotropy (due to e.\,g. Dresselhaus spin-orbit coupling) will also have a similar effect. We note, however, that all these mechanisms assume additional small factors that suppress the torques $b$,  $c$, and $d$ as compared to their bare values in Eq.~(\ref{bare}). 

\section{SOT in Dresselhaus model}
\label{sec:Dressel}

Let us now consider the Dresselhaus model of the form
\be
\label{Dressel}
H=\xi_{\bb{p}}+\alpha_\textrm{so} \, (\sigma_x p_x-\sigma_y p_y) - \bb{M}\cdot \bb{\sigma}+V(\bb{r}),
\e
where $\xi_{\bb{p}}$ is a function of the absolute value of the momentum. As before our symmetry analysis is valid for any $\xi_{\bb{p}}$ and any scalar disorder potential $V(\bb{r})$, while we use $\xi_{\bb{p}}=p^2/2m_e$ and Gaussian disorder for microscopic analysis. The model of Eq.~(\ref{Dressel}) can be transformed to the Rashba model by means of the unitary transformation
\be
H'=U\h H U= \xi_{\bb{p}}+\alpha_\textrm{so} \, (\bb{\sigma} \times \bb{p})_z  - \bb{M}'\cdot \bb{\sigma}+V(\bb{r}),
\e
where $U=(\sigma_x+\sigma_y)/\sqrt{2}$ and $\bb{M}'=(M_y,M_x,-M_z)$. Thus, the dynamics of the vector $\bb{m}'=\bb{M}'/J_\textrm{ex}$ is given by the Landau-Lifshitz-Gilbert equation (\ref{LLGmain}) with the SOT expressed by the Eq.~(\ref{torques}). By recasting the latter for vector $\bb{m}$ instead of $\bb{m}'$, one obtains the SOT for the Dresselhaus model in the form
\begin{align}
\bb{T}&= a\,\bb{m}\times \bb{E}_\textrm{D} + c\, \bb{m} \times (\bb{m}\times \hat{\bb{z}})\lt[(\hat{\bb{z}}\times \bb{m})\cdot \bb{E}_\textrm{D}\rt] \n \\
&+ b\,\bb{m}\times \lt(\bb{m}\times \bb{E}_\textrm{D}\rt) + d\, \bb{m} \times\hat{\bb{z}}\lt[(\hat{\bb{z}}\times \bb{m})\cdot \bb{E}_\textrm{D}\rt],
\label{TD}
\end{align}
where $\bb{E}_\textrm{D}=(E_x,-E_y)$. Note that the $x$ direction is specified in the case of the Dresselhaus type of spin-orbit interaction by the lattice orientation. In the full analogy with Eqs.~(\ref{torques}) and~(\ref{torques1}), one can also construct the torques that describe the response to the electric current rather than to electric field. 

Unlike the torque in the Rashba model (\ref{torques})-(\ref{Rashba}), the torque of Eq.~(\ref{TD}) does not depend on the direction of the vector $\hat{z}$. The coefficients $a$, $b$, $c$, and $d$ are, however, exastly the same as those defined for the Rashba model. Indeed, the substitution $m_x\to m_y$, $m_y\to m_x$ and $m_z\to -m_z$ cannot change the coefficients, because they depend only on $m_z^2$. Therefore, in the limit of Gaussian disorder treated within the self-consistent Born approximation for $\ep>E^*$, coefficient $a$ is finite but constant, while the other coefficients $b$, $c$ and $d$ are vanishing. 

\section{Skyrmion dynamics}
\label{sec:skyrm}

\subsection{Thiele equation}

Let us now apply the results of Eqs.~(\ref{torques}), (\ref{torques1}), and~(\ref{TD}) to analyze the motion of a Skyrmion by means of electric current. This problem can be considered along the lines of Refs.~\cite{Tretiakov2008,Clarke2008,Tveten13,Schuette14,Barker2016} by utilizing the automodel solution for the magnetization vector $\bb{m}=\bb{m}(\bb{r}-\bb{\nu}t)$, where $\bb{\nu}$ is the 2D velocity vector for a rigid Skyrmion spin-texture. The approach yields the so-called generalized Thiele equation for spin textures \cite{Thiele,Clarke2008}, which is derived in Appendix \ref{app:Thiele} in the following form
\be
\label{Thiele}
\big(Q\hat{\epsilon} -\hat{D}\big)\bb{\nu}=\bb{F},
\e
where $\hat{\epsilon}$ is the antisymmetric tensor with the components $\epsilon_{xy}=-\epsilon_{yx}=1$ and $\epsilon_{xx}=\epsilon_{yy}=0$. We have also introduced the quantities
\beml
\label{quantities}
\begin{align}
&Q=\frac{1}{4\pi} \int d^2\bb{r}\; \bb{m}\cdot [(\nabla_x \bb{m})\times (\nabla_y \bb{m})],\\
&D_{\alpha\beta}=\frac{\alpha_\textrm{G}}{4\pi}  \int d^2\bb{r}\; (\nabla_\alpha \bb{m})\cdot (\nabla_\beta \bb{m}),\\
&F_\alpha=\frac{\kappa}{4\pi} \int d^2\bb{r}\; (\nabla_\alpha \bb{m})\cdot \bb{s},
\label{F}
\end{align}
\eml
where the coefficient $Q$ is referred to as the topological charge, $\hat{D}$ is the dissipative tensor, and vector $\bb{F}$ is the generalized force that drives the spin texture. We restrict ourselves below to the case of azimuthally-symmetric Skyrmion that has the topological charge $Q=1$. 

An azimuthally-symmetric Skyrmion in a bilayer sample with perpendicular magnetic anisotropy (along $z$-direction) is parameterized by the magnetization vector $\bb{m}=(\cos\Phi \sin\theta, \sin\Phi \sin\theta,\cos\theta)$ with $\Phi=\varphi + \delta$, where $\rho$ and $\varphi$ are the polar coordinates with respect to the Skyrmion center. The function $\theta=\theta(\rho)$, which defines the Skyrmion profile, is material dependent, so we leave it unspecified. The phase $\delta$ is referred to as the helicity of the Skyrmion. In the case of Ne\'el Skyrmions, which are stabilized in the systems with strong interfacial Dzyaloshinskii-Moriya interaction typical for FM/HM bilayers, $\delta=0$. With the help of the above parameterization one finds from Eqs.~(\ref{quantities}) that the topological charge is, indeed, $Q=1$ \cite{Tretiakov2007} and that the dissipative tensor is diagonal $D_{\alpha\beta}=D\delta_{\alpha\beta}$, where
\be
D=\frac{\alpha_G}{4} \int_0^\infty \frac{d\rho}{\rho}\, \lt[\sin^2\theta+\lt(\rho\,\frac{\pa\theta}{\pa\rho}\rt)^2 \rt].
\e
The so-called Skyrmion Hall angle \cite{SOT exp Klaui} is defined by the ratio of velocity components. This ratio is found from Eq.~(\ref{Thiele}) as
\be
\frac{\nu_y}{\nu_x}=\frac{DF_y-F_x}{D F_x+F_y}.
\e
In the simple case $\bb{s}\propto \bb{H}_\textrm{eff}$, using Eq.~(\ref{F}) we find $\bb{F}=0$, thus confirming that the azimuthally-symmetric Skyrmion cannot be moved by an external field.

\subsection{Rashba model}

For Rashba model we have established the general expression of Eq.~(\ref{torques1}) for the electric-current driven SOT $\bb{T}_\textrm{J}=\kappa\,\bb{s}\times\bb{m}$, which corresponds to  
\begin{align}
\kappa\,\bb{s}=\,&-a_\textrm{J}\,\hat{\bb{z}}\times \bb{J}  -c_\textrm{J}\, \bb{m}\times \hat{\bb{z}}\;(\bb{m}\cdot \bb{J})\n\\
&- b_\textrm{J}\, \bb{m}\times [\hat{\bb{z}}\times \bb{J}] -  d_\textrm{J}\, \hat{\bb{z}}\,(\bb{m}\cdot \bb{J}),
\end{align} 
where the coefficients $a_\textrm{J}$, $b_\textrm{J}$, $c_\textrm{J}$, and $d_\textrm{J}$ may only depend on angle $\theta(\rho)$, because $m_z^2=\cos^2\theta$. Substituting this vector $\bb{s}$ into Eq.~(\ref{F}), we find the corresponding generalized force
\begin{align}
\label{forceR}
\bb{F}=\,&\frac{\hat{\bb{z}}\times\bb{J}_\delta}{4} \int_0^\infty\!\! d\rho\,\lt[\rho\frac{\pa a_\textrm{J}}{\pa \rho}+c_\textrm{J}\sin^2\theta\rt]\sin\theta\\
&+\frac{\bb{J}_\delta}{4} \int_0^\infty\!\! d\rho\, \lt[\frac{b_\textrm{J}}{2}\sin2\theta+(b_\textrm{J}+d_\textrm{J}\sin^2\theta)\rho\,\frac{\pa\theta}{\pa\rho}\rt],\n
\end{align}
where $\bb{J}_\delta= \bb{J}\cos\delta + (\bb{J}\times\hat{\bb{z}})\sin\delta$ is the current vector. Importantly, this vector is rotated on the angle given by helicity $\delta$, which changes from $\delta=0$ for the Ne\'el type of Skyrmions to $\delta=\pi/2$ for the Bloch type. All the integral coefficients in Eq.~(\ref{forceR}) depend, in general, on the Skyrmion profile $\theta(\rho)$. 

\subsection{Dresselhaus model}

Similar expression is readily obtained for the Dresselhaus model. The symmetry analysis expressed by the Eq.~(\ref{TD}) suggests that the electrical current driven SOT has a general form $\bb{T}_\textrm{J}=\kappa\,\bb{s}\times\bb{m}$ with
\begin{align}
\kappa\,\bb{s}=\,&-a_\textrm{J}\,\bb{J}_\textrm{D}  -c_\textrm{J}\, \bb{m}\times \hat{\bb{z}}\;\lt([\hat{\bb{z}}\times \bb{m}]\cdot \bb{J}_\textrm{D}\rt)\n\\
&- b_\textrm{J}\, \bb{m}\times \bb{J}_\textrm{D} -  d_\textrm{J}\, \hat{\bb{z}}\,\lt([\hat{\bb{z}}\times \bb{m}]\cdot \bb{J}_\textrm{D}\rt),
\end{align} 
where $\bb{J}_\textrm{D}=(J_x,-J_y)$. Substituting this expression into Eq.~(\ref{F}) we obtain for the Dresselhaus model
\begin{align}
\label{forceD}
\bb{F}=\,&\frac{\bb{J}^\textrm{D}_\delta}{4} \int_0^\infty\!\! d\rho\,\lt[\rho\frac{\pa a_\textrm{J}}{\pa \rho}+c_\textrm{J}\sin^2\theta\rt]\sin\theta\\
&- \frac{\hat{\bb{z}}\times\bb{J}^\textrm{D}_\delta}{4} \int_0^\infty\!\! d\rho\, \lt[\frac{b_\textrm{J}}{2}\sin2\theta+(b_\textrm{J}+d_\textrm{J}\sin^2\theta)\rho\,\frac{\pa\theta}{\pa\rho}\rt],\n
\end{align}
where $\bb{J}^\textrm{D}_\delta= \bb{J}_\textrm{D}\cos\delta + (\bb{J}_\textrm{D}\times\hat{\bb{z}})\sin\delta$. 

Thus, for an azimuthally-symmetric Skyrmion to be driven by SOT it is essential to have an angular dependence in the coefficient $a$ or finite values for the coefficients $b$, $c$ or $d$ irrespective of the nature of the spin-orbit interaction. 

A simple illustration of the results of Eqs.~(\ref{forceR}) and~(\ref{forceD}) is appropriate here. Suppose the coefficients $b_\textrm{J}$ and $d_\textrm{J}$ are negligibly small, as it must be the case for the limit of well-separated spin-split subbands. For the sake of definiteness let us consider the Ne\'el Skyrmion, which is characterized by $\delta=0$, and assume that the electric current is applied along $x$ direction. In this case we find $\bb{F}= A J \,\hat{\bb{y}}$ for the Rashba model and $\bb{F}= A J \,\hat{\bb{x}}$ for the Dresselhaus model, where the proportionality coefficient $A$ is set by the integral in Eq.~(\ref{forceR}) or Eq.~(\ref{forceD}), whereas vectors $\hat{\bb{x}}$ and $\hat{\bb{y}}$ are the unit vectors in $x$ and $y$ directions, respectively. The resulting Skyrmion Hall angle is then given by $\nu_y/\nu_x=D$ for the Rashba model, and by $\nu_y/\nu_x=-1/D$ for the Dresselhaus model. Both results are manifestly independent of the value of $A$. Meanwhile, for a Bloch Skyrmion (characterized by $\delta=\pi/2$) the results are reversed, namely the Hall angle for the Skyrmion motion is given by $\nu_y/\nu_x=-1/D$ for the Rashba model, and by $\nu_y/\nu_x=D$ for the Dresselhaus model. 

Remarkably, in the most general case, when all coefficients in Eqs.~(\ref{forceR}) and~(\ref{forceD}) are finite, one may see that the Hall angle for a Bloch Skyrmion with helicity $\delta=\pi/2$ is different from the one for a Skyrmion with helicity $\delta=-\pi/2$. Similarly, the Ne\'el Skyrmions with $\delta=0$ and $\delta=\pi$ move differently.  Yet some general relations may be established. For example, the motion of a Ne\'el  Skyrmion with $\delta=0$ in the Rashba model is identical to the motion of a Bloch Skyrmion with $\delta=-\pi/2$ in the Dresselhaus model if the current is applied along $x$ direction, and to the motion of a Bloch Skyrmion with $\delta=\pi/2$ in the Dresselhaus model if the current is applied along $y$ direction. These relations may be important in shedding light on the internal spin structure of Skyrmions in the experiments observing  Skyrmion dynamics \cite{SOT exp Klaui, Beach2016, Hoffmann16}.

Even though the presented microscopic calculations may not be used to predict absolute values of spin-orbit torques in real systems  (similarly as it is never possible to use model calculations to compute material conductivity), this model analysis captures important mutual relationships between different SOTs, which are universal beyond any specific model. Our results are also important for benchmarking of more general numerical methods based, for example, on the simulations of the corresponding Boltzmann equations for the magnetization and charge dynamics, which have yet to be accurately formulated.  

\section{Conclusions}
\label{sec:conclusion}

In conclusion, the symmetry of spin-orbit torques are identified for both electric-field and electric-current driven setups in two dimensions in the presence of spin-orbit interaction of either Rashba or Dresselhaus type. A general microscopic definition of the spin-orbit torques is given by relating them to susceptibility and conductivity tensors. The effect of SOTs on the motion of an azimuthally-symmetric Skyrmion is considered. The microscopic analysis of torques is performed for the generalized Bychkov-Rashba (or $s$-$d$-like) model with Gaussian scalar disorder within the self-consistent Born approximation. We demonstrate that the Skyrmion dynamics is completely suppressed in this model due to the exact cancelation of three out of four SOTs. Nevertheless, such an exact cancelation may be removed in the case of different density of states for two spin-split subbands, strong or spin-dependent disorder, or by taking into account the thermal fluctuations of Skyrmion shape. Those are examples of mechanisms that may help the spin-orbit torques to be effective for enabling Skyrmion dynamics. In addition, the spin-transfer torques in these spin-orbit systems, being sensitive to the gradients of magnetization, may prove to be more important for Skyrmion motion. The corresponding analysis will be published elsewhere. 

\acknowledgements

We are grateful to Artem Abanov, Geoffrey~Beach, Axel Hoffmann, Mathias Kl\"{a}ui, Alireza Qaiumzadeh, Koji Sato, and Oleg Tchernyshyov for helpful discussions. The work was supported by the Dutch Science Foundation NWO/FOM 13PR3118 and by the EU Network FP7-PEOPLE-2013-IRSES Grant No 612624 ``InterNoM''. O.\,A.\,T. acknowledges support by the Grants-in-Aid for Scientific Research (Grants No. 25800184, No. 25247056 and No. 15H01009) from MEXT, Japan and SpinNet.

\appendix

\section{Kubo formula for non-equilibrium spin polarization}
\label{app:Kubo}

In non-equilibrium approach to quantum mechanics (see Refs.~\cite{Kadanoff} and \cite{Rammer}), one defines the local quantum-mechanical average of electron-spin operator 
as
\be
\label{quant}
\bb{s}(\bb{r},t) = -\frac{i}{2} \tr_\sigma \bb{\sigma}\, \mathcal{G}^<(\bb{r},t;\bb{r},t),
\e
where the trace is taken only over the spin degree of freedom. The Green's function $\mathcal{G}^<(\bb{r}_1,t_1;\bb{r}_2,t_2)$ is a non-equilibrium Green's function that is conveniently represented as
\be
\mathcal{G}^<=\frac{1}{2}\lt(\mathcal{G}^K-\mathcal{G}^R+\mathcal{G}^A\rt)
\e
via the Keldysh $\mathcal{G}^K$, advanced $\mathcal{G}^A$, and retarded $\mathcal{G}^R$ Green's functions. The object $-i \mathcal{G}^<(\bb{r},t;\bb{r},t)$ is nothing but the density matrix of non-equilibrium quantum mechanics. In equilibrium all Green's functions have to be invariant with respect to time shifts, i.\,e. must depend only on the time difference $t-t'$. The translation invariance in real space is generally broken by disorder and can only be restored for disorder averaged quantities.

Let us use the Keldysh framework to define the linear response of the system to the external electric field $\bb{E}$. It is convenient to think of the electric field as a time-dependent perturbation to the model Hamiltonian
\be
H=H_0-\bb{j}\bb{A}(t),\qquad \bb{A}(t)=\frac{\bb{E}}{i\Omega}e^{-i\Omega t},
\e
where $\bb{A}(t)$ is the time-dependent vector potential, $\bb{j}$ is the current operator, and the {\em dc} limit $\Omega\to 0$ is assumed. It is also convenient to introduce a Keldysh space by organizing different Green's functions into the matrix
\be
\mathcal{G}=\bpm \mathcal{G}^R & \mathcal{G}^K\\ 0 & \mathcal{G}^A \epm,
\e
and define the Wigner transform $\mathcal{G}(\ep;t)$ with respect to time
\be
\mathcal{G}(t_1;t_2) =  \int \frac{d\ep}{2\pi} e^{-i\ep (t_1-t_2)} \mathcal{G}(\ep;t),\quad t=\frac{t_1+t_2}{2}, 
\e
where we suppress real-space indices, since they are largely irrelevant for the discussion below. The dependence on the absolute (physical) time $t$ is clearly absent in equilibrium. In this case, Green's function $\mathcal{G}^K$ is related to the functions $\mathcal{G}^R$ and $\mathcal{G}^A$ by means of the fluctuation-dissipation theorem,
\be
\label{FDT}
\mathcal{G}_0^K(\ep)=(\mathcal{G}_0^R(\ep)-\mathcal{G}_0^A(\ep))h_\ep,\quad h_\ep=\tanh\frac{\ep-\mu}{2T},
\e
where $\mu$ is the chemical potential and $T$ is the temperature. The relation (\ref{FDT}) does no longer hold in the presence of electric field. In the latter case the function $\mathcal{G}(\ep,t)$ acquires explicit $t$ dependence. In the first-order perturbation theory with respect to the electric field we, however, write $\mathcal{G} = \mathcal{G}_0 - \delta \mathcal{G}$, where 
\begin{align}
\label{basic0}
&\delta \mathcal{G}(t_1,\bb{r}_1;t_2,\bb{r}_2) \\
&= \int\! dt_3\!\int\! d\bb{r}_3\;  \mathcal{G}_0(t_1,\bb{r}_1;t_3,\bb{r}_3) \bb{j}_{\bb{r}_3}\bb{A}(t_3) \mathcal{G}_0(t_3,\bb{r}_3;t_2,\bb{r}_2),\n
\end{align}
and the matrix product in Keldysh space is assumed. The perturbation $\bb{j}\bb{A}$ is proportional to the unit matrix in Keldysh space, since we ignore quantum fluctuations of the electric field. 

Performing Wigner transform of Eq.~(\ref{basic0}) with respect to time, we arrive at the simple result:
\be
\label{basic1}
\delta \mathcal{G}(\ep,t)=\mathcal{G}_0(\ep+\Omega/2)\, \bb{j}\bb{A}(t)\,\mathcal{G}_0(\ep-\Omega/2),
\e
where the time convolution is absent but space convolution remains assumed. Equation~(\ref{basic1}) gives the Green's function components in Keldysh space
\begin{align}
&\delta \mathcal{G}^{R}(\ep,t)= \mathcal{G}^{R}_0(\ep+\Omega/2)\, \bb{j}\bb{A}(t)\,\mathcal{G}^{R}_0(\ep-\Omega/2),\\
&\delta \mathcal{G}^{A}(\ep,t)= \mathcal{G}^{A}_0(\ep+\Omega/2)\, \bb{j}\bb{A}(t)\,\mathcal{G}^{A}_0(\ep-\Omega/2),\\
&\delta \mathcal{G}^{K}(\ep,t)=
\mathcal{G}^{R}_0(\ep+\Omega/2)\, \bb{j}\bb{A}(t)\,\mathcal{G}^{R}_0(\ep-\Omega/2) h_{\ep-\Omega/2}\n\\
&\qquad-\mathcal{G}^{R}_0(\ep+\Omega/2)\, \bb{j}\bb{A}(t)\,\mathcal{G}^{A}_0(\ep-\Omega/2) h_{\ep-\Omega/2}\n\\
&\qquad+\mathcal{G}^{R}_0(\ep+\Omega/2)\, \bb{j}\bb{A}(t)\,\mathcal{G}^{A}_0(\ep-\Omega/2) h_{\ep+\Omega/2}\n\\
&\qquad-\mathcal{G}^{A}_0(\ep+\Omega/2)\, \bb{j}\bb{A}(t)\,\mathcal{G}^{R}_0(\ep-\Omega/2) h_{\ep+\Omega/2},
\end{align}
where we took advantage of Eq.~(\ref{FDT}). Collecting the results into $\delta\mathcal{G}^<=(\delta\mathcal{G}^K-\delta\mathcal{G}^R+\delta\mathcal{G}^A)/2$ and taking the limit $\Omega\to 0$ we obtain
\begin{align}
\delta \mathcal{G}^<(\ep,t)&=i\frac{\pa f}{\pa \ep}\lt[\mathcal{G}^{R} \bb{jE}\, \mathcal{G}^A\!-\!\tfrac{1}{2} \mathcal{G}^{R} \bb{jE}\, \mathcal{G}^R\!-\!\tfrac{1}{2} \mathcal{G}^{A} \bb{jE}\, \mathcal{G}^A\rt]\n\\
& +\frac{1}{2i}\lt(\mathcal{G}^{R} \bb{jE} \frac{\pa \mathcal{G}^R}{\pa \ep}-\frac{\pa \mathcal{G}^{R}}{\pa \ep} \bb{jE}\, \mathcal{G}^R \rt.\n\\
&\qquad \lt.- \mathcal{G}^{A} \bb{jE} \frac{\pa \mathcal{G}^A}{\pa \ep}+\frac{\pa \mathcal{G}^{A}}{\pa \ep} \bb{jE}\,\mathcal{G}^A \rt)f(\ep),
\label{Gless}
\end{align}
where we suppressed index $0$ and the argument $\ep$ on the Green's functions and introduced the Fermi distribution function $f(\ep)= (1-h_\ep)/2$. We have also omitted a term that is divergent in the limit $\Omega\to 0$ but does not contribute to the expression of Eq.~(\ref{quant}). We also note that the explicit dependence on physical time $t$ disappears in the zero-frequency limit.

To compute the average value of the electron spin operator in Eq.~(\ref{quant}) we have to take the result of Eq.~(\ref{Gless}) at coinciding space arguments. Since the Hamiltonian may contain explicit (though smooth) spacial dependence due to the dependence of the magnetization vector $\bb{m}$ on $\bb{r}$, we should employ another Wigner transform with respect to space arguments
\be
\mathcal{G}(\ep,t,\bb{r}_1,\bb{r}_2) =\int\frac{d^2\bb{p}}{(2\pi)^2}\,G(\ep,t;\bb{p},\bb{r})e^{i\bb{p}(\bb{r}_1-\bb{r}_2)},
\e
where $G(\ep,t;\bb{p},\bb{r})$ is now a smooth function of all its arguments. We, then, use the well-known property of Wigner transforms \cite{Rammer}
\be
\label{wigner}
(A\circ B)(\bb{r},p)=e^{\frac{i}{2}\lt(\bb{\nabla}^A_{\bb{r}} \bb{\nabla}^B_{\bb{p}}- \bb{\nabla}^A_{\bb{p}} \bb{\nabla}^B_{\bb{r}}\rt)} A(\bb{r},\bb{p}) B(\bb{r},\bb{p}),
\e
where $\circ$ stands for the convolution in real space. Equation (\ref{wigner}) sets out the gradient (adiabatic) expansion with respect to the slow variation of $\bb{m}$. In the leading (zeroth) order with respect to the gradient expansion we obtain the spin-orbit torques. The next order would give us spin-Hall related spin-transfer torques.  

In the leading order with respect to the magnetization gradients we simply replace the Wigner transform of the spacial convolution of Green's function with the product of Wigner transforms of the individual Green's functions in Eq.~\eqref{Gless}. As a result, we obtain the local linear response relation for the non-equilibrium spin density $\bb{s}=\hat{K}\bb{E}$, where 
$\hat{K}=\hat{K}^\textrm{I}+\hat{K}^\textrm{II}$. The tensors $\hat{K}^\textrm{I}$ and $\hat{K}^\textrm{II}$ are given by,
\begin{widetext}
\beml
\label{resss}
\begin{align}
\label{KI}
K^{\textrm{I}}_{\alpha\beta}= \,&\frac{1}{2}  \int \frac{d\ep}{2\pi} \int\frac{d^2\bb{p}}{(2\pi)^2}\, \lt(-\frac{\pa f}{\pa \ep}\rt) 
\tr_\sigma\lt\la \sigma_\alpha (G^R-G^A) j_\beta G^A - \sigma_\alpha G^R j_\beta (G^R-G^A) \rt\ra,\\
K^{\textrm{II}}_{\alpha\beta}=\, & \frac{1}{2}  \int \frac{d\ep}{2\pi} \int\frac{d^2\bb{p}}{(2\pi)^2}\, f(\ep) \tr_\sigma \lt\la
\sigma_\alpha G^{R} j_\beta \frac{\pa G^R}{\pa \ep}- \sigma_\alpha\frac{\pa G^{R}}{\pa \ep} j_\beta\, G^R - \sigma_\alpha G^{A} j_\beta \frac{\pa G^A}{\pa \ep}+\sigma_\alpha\frac{\pa G^{A}}{\pa \ep} j_\beta\,G^A \rt\ra,
\label{KII}
\end{align}
\eml
\end{widetext}
where the angular brackets indicate the averaging over disorder realizations. 

At zero temperature tensor $\hat{K}^\textrm{I}$ is clearly determined by the Green's functions at the Fermi energy, while the contribution $\hat{K}^\textrm{II}$ depends formally on all energies below the Fermi energy. The only contribution to the tensor $\hat{K}$ for $\ep > E^*$ is given by $\hat{K}^\textrm{I}$ that is provided at zero temperature by Eq.~(\ref{K}) of the main text.

\section{Relation between $\hat{K}_\textrm{J}$ and current-induced torques $\bb{T}_\textrm{J}$}
\label{app:KJ}

The relation between the tensor $\hat{K}_\textrm{J} =\hat{K}\hat{\sigma}^{-1}$ and the quantities $a_\textrm{J}$, $c_\textrm{J}$, $b_\textrm{J}$, and $d_\textrm{J}$, which define the current-induced torque $\bb{T}_\textrm{J}$ has been explained in the main text only in words. For the sake of completeness we quote here the corresponding formulas. Similarly to Eq.~(\ref{Ksym}), tensor $\hat{K}_\textrm{J}$ can be parameterized as 
\be
\label{KJsym}
\hat{K}_\textrm{J}= \frac{1}{\kappa}\bpm m_z \tilde{\kappa}_{xx} & \tilde{\kappa}_{xy}\\  \tilde{\kappa}_{yx} & m_z \tilde{\kappa}_{yy} \\  m_x \tilde{\kappa}_{zx} & m_x m_z \tilde{\kappa}_{zy}\epm,
\e
where we again refer to the normal coordinates in $xy$-plane with respect to the anisotropy (the $x$-axis is chosen along the in-plane magnetization component). The quantities $\tilde{\kappa}_{\alpha\beta}$ depend on $m_x^2=1-m_z^2$, hence all symmetry properties with respect to the inversion of the components are described by Eq.~(\ref{KJsym}).

Using that $\bb{T}_\textrm{J}=\kappa\, \bb{s}\times\bb{m}$, where $\bb{s}=\hat{K}_\textrm{J} \bb{J}$ we confirm the general ansatz of Eq.~(\ref{torques1}) and establish the following relations
\begin{align}
\label{symmetries2}
&a_\textrm{J}=\tilde{\kappa}_{xy}-m_x^2 \tilde{\kappa}_{zy},\qquad b_\textrm{J}=-\tilde{\kappa}_{yy},\n\\
&c_\textrm{J}=(\tilde{\kappa}_{xy}+\tilde{\kappa}_{yx})/m_x^2-\tilde{\kappa}_{zy},\\
&d_\textrm{J}=\tilde{\kappa}_{xx}+\tilde{\kappa}_{zx}+(\tilde{\kappa}_{yy}-\tilde{\kappa}_{xx})/m_x^2,\n
\end{align}
which connect spin-orbit torques of Eqs.~(\ref{torques1}) to electron spin susceptibilities described by $\hat{K}_\textrm{J}$. The results of Eqs.~(\ref{symmetries2}) can also be written explicitly as
\begin{align}
&a_\textrm{J}=K_{\textrm{J},xy}-\frac{M_x}{M_z} K_{\textrm{J},zy},\qquad b_\textrm{J}=-\frac{M}{M_z}K_{\textrm{J},yy},\n\\
&c_\textrm{J}=\frac{M^2}{M_x^2}\lt(K_{\textrm{J},xy}+K_{\textrm{J},yx}\rt)-\frac{M^2}{M_xM_z}K_{\textrm{J},zy},\\
&d_\textrm{J}=\frac{M}{M_z}K_{\textrm{J},xx}+\frac{M}{M_x}K_{\textrm{J},zx}+\frac{M^3}{M_xM_z^2}\lt(K_{\textrm{J},yy}-K_{\textrm{J},xx}\rt),\n
\end{align}
directly in terms of the components of tensor $\hat{K}_\textrm{J}$.

%%%%%%%%%%%%%%%%%%%%%%%%%%%%
%%%% fig:fig2sup
%%%%%%%%%%%%%%%%%%%%%%%%%%%%
\begin{figure*}[bt]
\includegraphics[width=2.0\columnwidth]{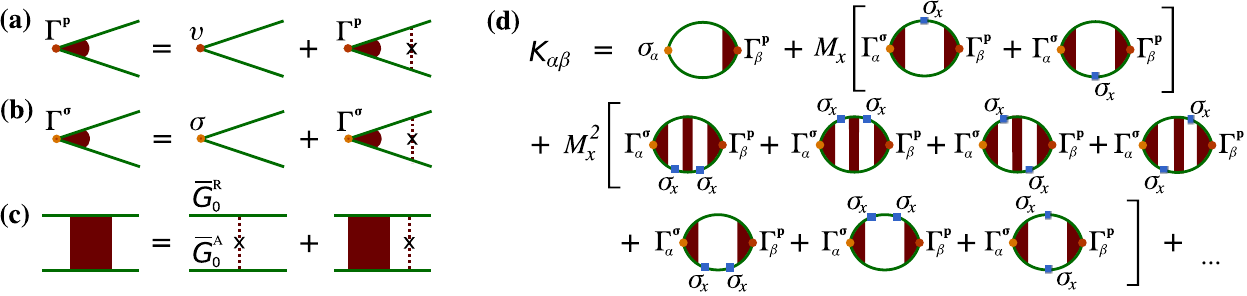}
\caption{(a) Diagrammatic representation of Eq.~(\ref{vertex}) on the vertex correction $\bb{\Gamma}^{\bb{p}}$ for the velocity operator $\bb{v}$; (b) The same but for the  the vertex correction $\bb{\Gamma}^{\bb{\sigma}}$; (c) Equation on the diffusion ladder; (d) Expansion of the disorder-averaged spin susceptibility tensor $\hat{K}$ in powers of in-plane magnetization component. The diffusion ladders and vertex corrections correspond to the so-called ``dressing'' of the bare diagrams depicted in Fig.~\ref{fig1}(b) in the main text.}
\label{fig2sup}
\end{figure*}
%%%%%%%%%%%%%%%%%%%%%%%%%%%%%

\section{Averaging over disorder}
\label{app:disorder}

The key building block of our diagrammatic analysis is the disorder-averaged Green's function $\bar{G}_0$ for the isotropic Bychkov-Rashba model
\be
H_0=\frac{p^2}{2m_e} +\alpha_\textrm{R} \, (\bb{\sigma} \times \bb{p})_z - M_z \sigma_z + V(\bb{r}),
\e
where the scalar Gaussian disorder potential $V(\bb{r})$ is characterized by the correlator $\la V(\bb{r})V(\bb{r}')\ra=\alpha_\textrm{D}\delta(\bb{r}-\bb{r}')$, $\alpha_\textrm{d}=(m_e\tau)^{-1}$.  To make our notations more economic we use $G=\bar{G}_0$ in this section. The averaged retarded Green's function in the Born approximation is characterized by the self-energy $\Sigma^R$, which is particularly simple in the upper band, $\ep>M_z$. In this case the self-energy lacks a matrix structure and is simply given by  $\Sigma^R_0=-i\gamma=-i/2\tau$. The resulting averaged Green's function in the Born approximation $G^R_{\bb{p}}=[\ep-H_0-\Sigma^R_0]^{-1}$ can also be written as
\be
G^R_{\bb{p}}==\frac{\ep+i\gamma-\xi+\sqrt{2\xi \Delta}\sigma_\phi-M_z\sigma_z}{(\xi-x_+)(\xi-x_-)},
\e
where $\bb{p}=p(\cos\phi,\sin\phi)$, $\sigma_\phi=\sigma_x\sin\phi-\sigma_y\cos\phi$, $\Delta=m_e\alpha_\textrm{R}^2$, $\xi=p^2/2m_e$, and
\be
\label{xnote}
x_{\pm}=\ep+i\gamma+\Delta\mp \sqrt{\lambda^2+2i\gamma\Delta},
\e
with the parameter $\lambda=\sqrt{\Delta^2+2\ep\Delta+M_z^2}$. For more details see also Ref.~\cite{Ado2}.

Calculation of vertex corrections is facilitated by the following integrals
\begin{align}
\label{aux}
&\alpha_\textrm{d}\int \frac{d^2\bb{p}}{(2\pi)^2} G^A_{\bb{p}} \sigma_x G^R_{\bb{p}} =\frac{\ep\Delta}{M_z^2+2\ep\Delta}\sigma_x-\frac{M_z\gamma}{M_z^2+2\ep\Delta}\sigma_y,\n\\
&\alpha_\textrm{d}\int \frac{d^2\bb{p}}{(2\pi)^2} G^A_{\bb{p}} \sigma_y G^R_{\bb{p}} =\frac{\ep\Delta}{M_z^2+2\ep\Delta}\sigma_y+\frac{M_z\gamma}{M_z^2+2\ep\Delta}\sigma_x,\n\\
&\alpha_\textrm{d}\int \frac{d^2\bb{p}}{(2\pi)^2} G^A_{\bb{p}} \lt(\frac{p_x}{\alpha_\textrm{R}m_e}\rt) G^R_{\bb{p}}= \sigma_y,\\
&\alpha_\textrm{d}\int \frac{d^2\bb{p}}{(2\pi)^2} G^A_{\bb{p}} \lt(\frac{p_y}{\alpha_\textrm{R}m_e}\rt) G^R_{\bb{p}}= -\sigma_x,\n\\
&\alpha_\textrm{d}\int \frac{d^2\bb{p}}{(2\pi)^2} G^A_{\bb{p}} \sigma_z G^R_{\bb{p}} =\frac{M_z^2}{M_z^2+2\ep\Delta}\sigma_z,\n
\end{align}
that are taken up here to the second order in $\gamma$. This precision is sufficient to compute the leading and sub-leading terms in $\hat{K}$ and $\hat{\sigma}$ with respect to the scattering time $\tau$ in the non-crossing (self-consistent Born) approximation. Vertex corrections yield the equations represented diagrammatically in Fig.~\ref{fig2sup}(a-b),
\beml
\label{vertex}
\begin{align}
&\bb{\Gamma}^{\bb{p}}=\bb{v}+\alpha_\textrm{d}\int \frac{d^2\bb{p}}{(2\pi)^2} G^A_{\bb{p}} \bb{\Gamma}^{\bb{p}} G^R_{\bb{p}},\\
&\bb{\Gamma}^{\bb{\sigma}}=\bb{\sigma}+\alpha_\textrm{d}\int \frac{d^2\bb{p}}{(2\pi)^2} G^A_{\bb{p}} \bb{\Gamma}^{\bb{\sigma}} G^R_{\bb{p}},
\end{align}
\eml
which are solved with the help of Eqs.~(\ref{aux}) as
\beml
\begin{align}
&\bb{\Gamma}^{\bb{p}}=\frac{\bb{p}}{m},\qquad \bb{\Gamma}_{z}^{\bb{\sigma}}=\lt(1+\frac{M_z^2}{2\ep\Delta}\rt)\sigma_z,\\  
&\bb{\Gamma}_{x}^{\bb{\sigma}}=\lt(1+\frac{\ep\Delta}{M_z^2+\ep\Delta}\rt)\lt(\sigma_x-\frac{M_z\gamma}{M_z^2+\ep\Delta}\sigma_y\rt),\\
&\bb{\Gamma}_{y}^{\bb{\sigma}}=\lt(1+\frac{\ep\Delta}{M_z^2+\ep\Delta}\rt)\lt(\sigma_y+\frac{M_z\gamma}{M_z^2+\ep\Delta}\sigma_x\rt).
\end{align}
\eml

Using the relations of Eq.~(\ref{aux}), we now reproduce the result for the fully dressed $\hat{K}$-tensor at $M_x=0$ 
\beml
\label{trivial}
\begin{align}
K^{(0)}_{\alpha x}=\frac{e\alpha_\textrm{R}}{4\pi\alpha_\textrm{D}}  \tr \sigma_\alpha \sigma_y = \frac{e \tau \alpha_\textrm{so} m_e}{2\pi} \delta_{\alpha y},\\
K^{(0)}_{\alpha y}=-\frac{e\alpha_\textrm{R}}{4\pi\alpha_\textrm{D}}  \tr \sigma_\alpha \sigma_x = -\frac{e \tau \alpha_\textrm{so} m_e}{2\pi} \delta_{\alpha x},
\end{align}
\eml
where $\delta_{\alpha\beta}$ is the Kronecker delta. This result is just a particular case of Eq.~(\ref{dressed}) of the main text for $M_x=0$. The calculation of conductivity is fully analogous. The bare tensor in the limit $M_x=0$ is, however, much more complex. It is given by the components
\begin{align}
&K^{\textrm{bare},(0)}_{\alpha x}=\frac{e}{4\pi}\tr \sigma_\alpha \int \frac{d^2\bb{p}}{(2\pi)^2} G^R_{\bb{p}} \lt(\frac{p_x}{m_e}-\alpha_\textrm{R}\sigma_y\rt) G^A_{\bb{p}} \n\\
&=\frac{e \alpha_\textrm{R}m_e}{4\pi} \lt[2\tau \frac{M_z^2+\ep\Delta}{M_z^2+2\ep\Delta}\delta_{\alpha y}-\frac{M_z}{M_z^2+2\ep\Delta}\delta_{\alpha x}\rt],\n\\
&K^{\textrm{bare},(0)}_{\alpha y}=\frac{e}{4\pi}\tr \sigma_\alpha \int \frac{d^2\bb{p}}{(2\pi)^2} G^R_{\bb{p}} \lt(\frac{p_y}{m_e}+\alpha_\textrm{R}\sigma_x\rt) G^A_{\bb{p}}\n\\
&=-\frac{e \alpha_\textrm{R}m_e}{4\pi} \lt[2\tau \frac{M_z^2+\ep\Delta}{M_z^2+2\ep\Delta}\delta_{\alpha x}+\frac{M_z}{M_z^2+2\ep\Delta}\delta_{\alpha y}\rt],\n
\end{align}
where we again took advantage of Eqs.~(\ref{aux}).

%%%%%%%%%%%%%%%%%%%%%%%%%%%%
%%%% fig:fig3sup
%%%%%%%%%%%%%%%%%%%%%%%%%%%%
\begin{figure*}[bt]
\includegraphics[width=1.4\columnwidth]{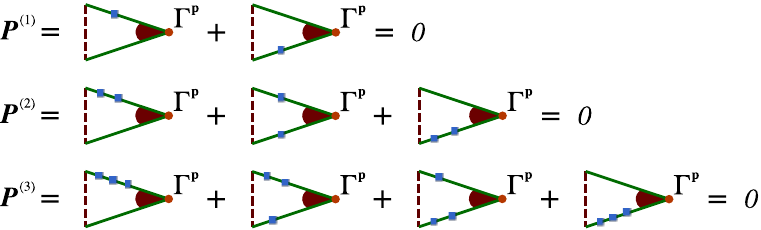}
\caption{Diagrammatic representation of exact cancelation of $\bb{P}^{(n)}$ as expressed by Eq.~(\ref{Pcancel}) in the model with gaussian spin-independent disorder.  Squares denote the inclusions of $\sigma_x$ matrices due to perturbation theory construction with respect to the term $M_x\sigma_x$. The cancelation persists in all orders of the perturbation theory. As a result, only the first diagram in Fig.~\ref{fig2sup}(d) for tensor $\hat{K}$ is finite. Consequently, tensor $K$ in non-crossing approximation does not depend on magnetization vector $\bb{m}$ (see Eq.~(\ref{dressed}) of the main text).}
\label{fig3sup}
\end{figure*}
%%%%%%%%%%%%%%%%%%%%%%%%%%%%%

Let us now outline the computation of the bare tensor $\hat{K}^\textrm{bare}$ in the expansion over the anisotropy as it is illustrated in Fig.~\ref{fig1} of the main text, $\hat{K}=\hat{K}^{(0)}+M_x \hat{K}^{(1)}+M_x^2 \hat{K}^{(2)}+\dots$. 

Such an expansion can be routinely computed using Mathematica package by noting that the integration over the variable $\xi$ can be extended to the entire real axis as far as we are interested in the leading and sub-leading orders with respect to $\gamma$ (we remind that the non-crossing approximation is not a consistent approximation in the sub-leading (zeroth) order with respect to $\gamma$  \cite{Ado2}). For the first two terms we obtain
\begin{align}
&K^{\textrm{bare},(1)}_{\alpha\beta}=-\frac{e}{4\pi} \int \frac{d^2\bb{p}}{(2\pi)^2}
\tr \lt[\sigma_\alpha G^R_{\bb{p}} \sigma_x G^R_{\bb{p}} v_\beta G^A_{\bb{p}} \rt.\n\\
&\qquad\qquad\qquad\qquad\lt.+\sigma_\alpha G^R_{\bb{p}}v_\beta G^A_{\bb{p}} \sigma_x G^A_{\bb{p}} \rt],\n\\
&K^{\textrm{bare},(2)}_{\alpha\beta}=\frac{e}{4\pi}  \int \frac{d^2\bb{p}}{(2\pi)^2} 
\tr \lt[\sigma_\alpha G^R_{\bb{p}} \sigma_x G^R_{\bb{p}} v_\beta G^A_{\bb{p}} \sigma_x G^A_{\bb{p}} \rt.\n\\
&\qquad\lt.+\sigma_\alpha \lt(G^R_{\bb{p}}\sigma_x\rt)^2 G^R_{\bb{p}}v_\beta  G^A_{\bb{p}} + \sigma_\alpha  G^R_{\bb{p}}v_\beta G^A_{\bb{p}}  \lt(\sigma_xG^A_{\bb{p}} \rt)^2\rt].\n
\end{align}
Explicit calculation gives the bare tensors
\beml
\label{baresup}
\begin{align}
\hat{K}^{\textrm{bare},(1)}= &\;\frac{e \alpha_\textrm{so} m_eM_z^2}{4\pi(M_z^2+2\ep\Delta)^2}
\bpm 0&0\\0&0\\ 1& M_z/\gamma\epm,\\
\hat{K}^{\textrm{bare},(2)} = &\; -\frac{e \alpha_\textrm{so} m_e M_z}{4\pi(M_z^2+2\ep\Delta)^3}\n\\
&\times
\bpm 2\ep\Delta-M_z^2 &(M_z+\ep\Delta)M_z/\gamma \\ -\ep\Delta M_z/\gamma &  2\ep\Delta-M_z^2 \\ 0& 0 \epm.
\end{align}
\eml
Similar analysis can be performed for the conductivity. The results are summarized in Eq.~(\ref{bare}) of the main text up to the terms of the order of $M_x^3$.

Computation of the fully dressed tensor is much more simple. It can be computed without resorting to the bare tensors (\ref{baresup}). Indeed, from the very beginning we can use the fact that the dressed current operator is purely kinematic $\bb{\Gamma}^{\bb{p}}=\bb{p}/m_e$. To prove that the full tensor $\hat{K}$ and $\hat{\sigma}$ do not depend on the direction of magnetization it sufficient to analyze the tensors
\be
\bb{P}^{(n)}= \s_{m=0}^n \int \frac{d^2\bb{p}}{(2\pi)^2} \lt[G^R_{\bb{p}} \sigma_x\rt]^{m}\!G^R_{\bb{p}}\, \bb{\Gamma}^{\bb{p}}\, G^A_{\bb{p}} \lt[\sigma_x G^A_{\bb{p}}\rt]^{n-m},\n
\e
that are shown schematically in Fig.~\ref{fig3sup}. The direct computation gives 
\be
\label{Pcancel}
\bb{P}^{(n)}=0,
\e
for any value of $n$. In fact we have checked the identity (\ref{Pcancel}) analytically up to $n=7$ but did not find a rigorous general proof. In doing this calculation, it is important not to expand the square roots in Eqs.~(\ref{xnote}) over $\gamma$. The identity (\ref{Pcancel}) assumes the following relation 
\be
\int \frac{d^2\bb{p}}{(2\pi)^2} \bar{G}^R_{\bb{p}}\bb{p}\,  \bar{G}^A_{\bb{p}} = \int \frac{d^2\bb{p}}{(2\pi)^2} G^R_{\bb{p}}\bb{p}\,  G^A_{\bb{p}}, 
\e
where the Green's function
\be
\bar{G}^R_{\bb{p}} = \lt[\ep+i\gamma -\xi_p-\alpha_\textrm{R} (\bb{\sigma}\times\bb{p})_z+M_x \sigma_x+M_z\sigma_z\rt]^{-1},\nonumber\\
\e
stands for the averaged Green's function (in the Born approximation for $\ep>E^*$) for the anisotropic model. 

It follows immediately from Eq.~(\ref{Pcancel}) that all diagrams involving one or more $\sigma_x$ matrix are identically zero after dressing the current vertex. Therefore, the dressed tensors $\hat{K}$ and $\hat{\sigma}$ are simply identical to those for $M_x=0$. The latter, in turn, do not depend on $M_z$. Thus, we conclude that in the model with scalar gaussian disorder both $\hat{K}$ and $\hat{\sigma}$ tensors do not depend on $\bb{M}$ in the non-crossing approximation. 

%We know that this property does not hold beyond the non-crossing approximation due to the scattering on rare impurity configurations \cite{Ado2sup}.

\section{Generalized Thiele equation}
\label{app:Thiele}

In this section we derive Eq.~(\ref{Thiele}) of the main text. We start from the LLG equation in the form
\be
\label{LLG}
\dot{\bb{m}}=\bb{f}\times\bb{m}+\alpha_G\,\bb{m}\times\dot{\bb{m}},
\vspace*{6pt}
\e
where $\bb{f}=\gamma \bb{H}_{\textrm{eff}} +\kappa \bb{s}$ (cf. Eq.~(\ref{ffield})), and apply the automodel ansatz $\bb{m}=\bb{m}(\bb{r}-\bb{\nu}t)$, where $\bb{\nu}$ is the velocity of the spin texture, and hence $\dot{\bb{m}}=-\nu_\beta \nabla_\beta \bb{m}$. Here and below the summation over the repeated index $\beta$ is assumed. Therefore, we can rewrite Eq.~(\ref{LLG}) in the form
\be
\lt(\nabla_\beta \bb{m}-\alpha_G\,\bb{m}\times \nabla_\beta \bb{m}\rt)\nu_\beta +\bb{f}\times\bb{m}=0.
\e
By taking the vector product of this equation with $\bb{m}$ and using the identity $\bb{m} \cdot \nabla_\beta \bb{m} =0$, we find
\be
\lt(\bb{m}\times \nabla_\beta \bb{m}+\alpha_G\,\nabla_\beta \bb{m}\rt)\nu_\beta+\bb{f} -\kappa\,\bb{m}(\bb{m}\cdot\bb{s})=0.\n
\e   
We now take the scalar product of this equation with the vector $\nabla_\alpha\bb{m}$ to obtain
\begin{align}
\big(\bb{m} \cdot [(\nabla_\alpha \bb{m})\times(\nabla_\beta \bb{m})] -&\,\alpha_G\,(\nabla_\alpha \bb{m})\cdot(\nabla_\beta \bb{m})\big)\nu_\beta \n\\
&=\bb{f}\cdot (\nabla_\alpha \bb{m}).
\end{align}
Integrating the last equation over the space and dividing by $4\pi$ we reproduce the result of Eq.~(\ref{Thiele}) of the main text.


\begin{thebibliography}{99}

\bibitem{Rashba-SOI}
C.\,R.~Ast, J.~Henk, A.~Ernst, L.~Moreschini, M.\,C.~Falub, D.~Pacil\'{e}, P.~Bruno, K.~Kern, and M.~Grioni, 
Phys.\ Rev.\ Lett.  \textbf{98}, 186807 (2007).

\bibitem{Hoffmann13}%Spin Hall effects in metals
A.~Hoffmann, IEEE Trans. Magn. \textbf{49}, 5172 (2013).

\bibitem{review-Rashba1}
D.~Bercioux and P.~Lucignano, Rep.\ Prog.\ Phys. \textbf{78}, 106001 (2015).

\bibitem{review-Rashba2}%New perspectives for Rashba spin-orbit coupling
A.~Manchon, H.\,C.~Koo,	J.~Nitta, S.\,M.~Frolov, and R.\,A.~Duine,  
Nature Materials \textbf{14}, 871 (2015).

\bibitem{Sinova15}%Spin Hall effects
J.~Sinova, S.\,O.~Valenzuela, J.~Wunderlich, C.\,H.~Back, and T.~Jungwirth,
Rev.\ Mod.\ Phys. \textbf{87}, 1213 (2015); 

\bibitem{Garello-switching} 
K.~Garello, C.\,O.~Avci, I.\,M.~Miron, M.~Baumgartner, A.~Ghosh, S.~Auffret, O.~Boulle, G.~Gaudin, and P.~Gambardella, 
Appl.\ Phys.\ Lett. \textbf{105}, 212402 (2014).

\bibitem{SOT1} 
B.\,A.~Bernevig and O.~Vafek, 
Phys.\ Rev.\ B \textbf{72}, 033203 (2005).

\bibitem{SOT2} 
A.~Manchon and S.~Zhang, 
Phys.\ Rev.\ B \textbf{78}, 212405 (2008); \textbf{79}, 094422 (2009).

\bibitem{SOT3} 
A.~Matos-Abiague and R.\,L.~Rodr\'{\i}guez-Su\'{a}rez, Phys.\ Rev.\ B \textbf{80}, 094424 (2009).

\bibitem{SOT4} 
I.~Garate, A.\,H.~MacDonald, Phys.\ Rev.\ B \textbf{80}, 134403 (2009).

\bibitem{anti-damping SOT1} 
X.~Wang and A.~Manchon, 
Phys.\ Rev.\ Lett. \textbf{108}, 117201 (2012).

\bibitem{anti-damping SOT2} 
K.\,-W.~Kim, S.\,-M.~Seo, J.~Ryu, K.\,-J.~Lee, and H.\,-W.~Lee, 
Phys.\ Rev.\ B \textbf{85}, 180404(R) (2012).

\bibitem{anti-damping SOT-QKE}  
D.\,A.~Pesin and A.\,H.~MacDonald, 
Phys.\ Rev.\ B \textbf{86}, 014416 (2012).

\bibitem{SOT AFM1} 
J.~\v{Z}elezn\'{y}, H.~Gao, K.~V\'{y}born\'{y}, J.~Zemen, J.~Ma\v{s}ek, A.~Manchon, J.~Wunderlich, J.~Sinova, and T.~Jungwirth, 
Phys.\ Rev.\ Lett. \textbf{113}, 157201 (2014).

\bibitem{SOT_Manchon2015} 
H. Li, H. Gao, L. P. Z\^arbo, K.~V\'{y}born\'{y}, X. Wang, I. Garate, F. Do\ifmmode \check{g}\else \v{g}\fi{}an, A. \ifmmode \check{C}\else \v{C}\fi{}ejchan, J.~Sinova, T.~Jungwirth, and A.~Manchon,
Phys.\ Rev.\ B \textbf{91}, 134402 (2015).

\bibitem{SOT AFM2}%First all-antiferromagnetic memory device could get digital data storage in a spin
P.~Wadley, B.~Howells, J.~\v{Z}elezn\'{y}, C.~Andrews, V.~Hills, R.\,P.~Campion, V.~Nov\'{a}k, F.~Freimuth, Y.~Mokrousov, A.\,W.~Rushforth, K.\,W.~Edmonds, B.\,L.~Gallagher, T.~Jungwirth, arXiv: 1503.03765 (2015).

\bibitem{Emori13}%Current-driven dynamics of chiral ferromagnetic domain walls.
S.~Emori, U.~Bauer, S.\,M.~Ahn, E.~Martinez, G.\,S.~Beach,  
Nature Materials \textbf{12}, 611 (2013).

\bibitem{Ryu13}%Chiral spin torque at magnetic domain walls
K.\,S.~Ryu, L.~Thomas, S.\,H.~Yang, S.~Parkin, 
Nature Nanotechnology \textbf{8}, 527 (2013).

\bibitem{SOT exp}%Evidence for reversible control of magnetization in a ferromagnetic material via spin-orbit magnetic field
A.~Chernyshov, M.~Overby, X.~Liu, J.\,K.~Furdyna, Y.~Lyanda-Geller, and L.\,P.~Rokhinson, 
Nature Physics \textbf{5}, 656 (2009).

\bibitem{SOT DW Miron1}
I.\,M.~Miron, T.~Moore, H.~Szambolics, L.\,D.~Buda-Prejbeanu, S.~Auffret, B.~Rodmacq, S.~Pizzini, J.~Vogel, M.~Bonfim, A.~Schuhl, and G.~Gaudin, 
Nature Materials \textbf{10}, 419 (2011).
 
\bibitem{SOT DW Miron2}%Matching domain-wall configuration and spin-orbit torques for efficient domain-wall motion
A.\,V.~Khvalkovskiy, V.~Cros, D.~Apalkov, V.~Nikitin, M.~Krounbi, K.\, A.~Zvezdin, A.~Anane, J.~Grollier, A.~Fert, 
Phys.\ Rev.\ B \textbf{87}, 020402(R)
(2013).

\bibitem{SOT exp Klaui}
K. Litzius, I. Lemesh, B.~Kr\"{u}ger, P.~Bassirian, L.~Caretta, K.~Richter, F.~B\"{u}ttner, K.~Sato, O.\,A.~Tretiakov, J.~F\"{o}rster, R.\,M.~Reeve, M.~Weigand, I.~Bykova, H.~Stoll, G.~Sch\"{u}tz, G. S. D. Beach, and M.~Kl\"{a}ui, %Skyrmion Hall Effect Revealed by Direct Time-Resolved X-Ray Microscopy,
Nature Physics, doi:10.1083/nphys4000 (2016).

\bibitem{SOT exp Miron1} 
I.\,M.~Miron, G.~Gaudin, S.~Auffret, B.~Rodmacq, A.~Schuhl, S.~Pizzini, J.~Vogel, and P.~Gambardella, 
Nature Materials \textbf{9}, 230 (2010).

\bibitem{SOT exp Miron2}%Perpendicular switching of a single ferromagnetic layer induced by in-plane current injection. 
I.\,M.~Miron, K.~Garello, G.~Gaudin, P.\,-J.~Zermatten, M.\,V.~Costache, S.~Auffret, S.~Bandiera, B.~Rodmacq, A.~Schuhl, and P.~Gambardella, 
Nature \textbf{476}, 189 (2011).

\bibitem{SOT exp Miron3} 
K.~Garello, I.\,M.~Miron, C.\,O.~Avci, F.\,Freimuth, Y.~Mokrousov, S.~Bl\"{u}gel, S.~Auffret, O.~Boulle, G.~Gaudin, and P.~Gambardella, 
Nature Nanotechnology \textbf{8}, 587 (2013).

\bibitem{Liu12}%Spin-Torque Switching with the Giant Spin Hall Effect of Tantalum
L.\,Q.~Liu, C.\,F.~Pai, Y.~Li, H.\,W.~Tseng, D.\,C.~Ralph, R.\,A.~Buhrman,  
Science \textbf{336}, 555 (2012).

\bibitem{co-pt}%Orbital chirality and Rashba interaction in magnetic bands
J.\,-H.~Park, C.\,H.~Kim, H.\,-W.~Lee, and J.\,H.~Han, Phys.\ Rev.\ B \textbf{87}, 041301(R) (2013).

\bibitem{Tomasello2014}%A strategy for the design of Skyrmion racetrack memories
R.~Tomasello, E.~Martinez, R.~Zivieri, L.~Torres, M.~Carpentieri, and G.~Finocchio, Scientific Reports \textbf{4}, 6784 (2014).

\bibitem{Linder13}%Chirality Sensitive Domain Wall Motion in Spin-Orbit Coupled Ferromagnets
J.~Linder, Phys.\ Rev.\ B \textbf{87}, 054434 (2013)

\bibitem{SOTreview1}
P.~Gambardella and I.\,M.~Miron, 
Phil.\ Trans.\ R.\ Soc.\ A \textbf{369}, 3175 (2011).

\bibitem{SOTreview2}
A.~Brataas and K.\,M.\,D.~Hals, 
Nature Nanotechnology \textbf{9}, 86 (2014).

\bibitem{SOTreview3}
K.\,M.\,D.~Hals and A.~Brataas, 
Phys.\ Rev.\ B \textbf{88}, 085423 (2013).

\bibitem{Hoffmann15}%Blowing Magnetic Skyrmion Bubbles
W.~Jiang, P.~Upadhyaya, W.~Zhang, G.~Yu, M.\,B.~Jungfleisch, F.\,Y.~Fradin, J.\,E.~Pearson, Y.~Tserkovnyak, K.\,L.~Wang, O.~Heinonen, S.\,G.\,E.\,te~Velthuis, A.~Hoffmann, Science, \textbf{349}, 283 (2015).

\bibitem{Hoffmann16}%Direct Observation of the Skyrmion Hall Effect
W.~Jiang, X.~Zhang, G.~Yu, W.~Zhang, M.\,B.~Jungfleisch, J.\,E.~Pearson,  O.~Heinonen, K.\,L.~Wang, Y.~Zhou, A.~Hoffmann, and S.\,G.\,E.\,te~Velthuis, 
arXiv:1603.07393 (2016);

\bibitem{Beach2016} %Observation of room-temperature magnetic Skyrmions and their current-driven dynamics in ultrathin metallic ferromagnets
S. Woo, K.~Litzius, B.~Kr\"{u}ger, M.-Y.~Im, L.~Caretta, K.~Richter, M.~Mann, A.~Krone, R.\,M.~Reeve, M.~Weigand, P.~Agrawal,	I.~Lemesh, M.-A.~Mawass, P.~Fischer, M.~Kl\"{a}ui, and G.\,S.\,D.~Beach,
Nature Materials, doi:10.1038/nmat4593  (2016).

\bibitem{Jonietz10}
F.~Jonietz, S.~Muhlbauer, C.~Pfleiderer, A.~Neubauer, W.~Munzer, A.~Bauer, T.~Adams, R.~Georgii, P.~Boni, R.\,A.~Duine, K.~Everschor, M.~Garst and A.~Rosch,
Science \textbf{330}, 1648 (2010).

\bibitem{Bader15}
A.~Hoffmann and S.\,D.~Bader, 
Phys.\ Rev.\ App. \textbf{4}, 047001 (2015).

\bibitem{SHEsinova}
J.~Sinova, S.\,O.~Valenzuela, J.~Wunderlich, C.\,H.~Back, and T.~Jungwirth, Rev.\ Mod.\ Phys. \textbf{87}, 1213 (2015). 

\bibitem{STTreview1} 
N.~Locatelli, V.~Cros and J.~Grollier, Nat.\ Mater. \textbf{13}, 11 (2013).

\bibitem{STTreview2} 
A.~Brataas, A.\,D.~Kent, and H.~Ohno, Nat.\ Mater. \textbf{11}, 372 (2012).

\bibitem{STTreview3} 
D.\,C.~Ralph and M.\,D.~Stiles, J.\ Magn,\ Magn,\ Matter. \textbf{320}, 1190 (2008).

\bibitem{Bychkov84}
Y.\,A.~Bychkov and E.\,I.~Rashba, J.\ Phys.\ C \textbf{17}, 6039 (1984).

\bibitem{Streda}
P.~St\v{r}eda,
J.\ Phys.\ C: Solid State Phys.\ \textbf{15}, L717 (1982).

\bibitem{Bijl12}%Current-induced torques in textured Rashba ferromagnets
E.~van der Bijl and R.\,A.~Duine, 
Phys.\ Rev.\ B \textbf{86}, 094406 (2012) 

\bibitem{Hals13}%Phenomenology of current-induced spin-orbit torques
K.\,M.\,D.~Hals and A.~Brataas, 
Phys.\ Rev.\ B \textbf{88},  085423 (2013).

\bibitem{Lee15}%Angular dependence of spin-orbit spin-transfer torques
K.-S.~Lee, D.~Go, A.~Manchon, P.\,M.~Haney, M.\,D.~Stiles, H.-W.~Lee, and K.-J.~Lee,
Phys.\ Rev.\ B \textbf{91}, 144401 (2015).

\bibitem{footnote}
One can estimate the threshold energy for the upper band as $E^*\leq m_e \alpha_\textrm{R}^2/2+ \sqrt{M_z^2+(|M_x|- m_e \alpha_\textrm{R}^2)^2}$.

\bibitem{Ado2}
I.\,A.~Ado, I.\,A.~Dmitriev, P.\,M.~Ostrovsky, and M.~Titov, 
Phys.\ Rev.\ Lett. \textbf{117}, 046601 (2016)

\bibitem{Berry}
M.\,V. Berry, Proc.\ R.\ Soc.\ Lond. \textbf{392}, 45 (1984).

\bibitem{Ado1}
I.\,A.~Ado, I.\,A.~Dmitriev, P.\,M.~Ostrovsky, and M.~Titov, EPL \textbf{111}, 37004 (2015).

\bibitem{Edelstein effect1}
V.\,M.~Edelstein, Solid State Commun. \textbf{73}, 233 (1990).

\bibitem{Edelstein effect2}
V.\,M.~Edelstein, Phys.\ Rev.\ Lett. \textbf{80}, 5766 (1998).

\bibitem{Edelstein effect3}
J.\,-i.~Inoue, G.\,E.\,W.~Bauer, and L.\,W.~Molenkamp, Phys.\ Rev.\ B \textbf{67}, 033104 (2003).

\bibitem{Inoue2004}%Suppression of the persistent spin Hall current by defect scattering, 
J.-i.~Inoue, G.\,E.\,W.~Bauer, and L.\,W.~Molenkamp, Phys.\ Rev.\ B \textbf{70}, 041303(R) (2004).

\bibitem{Ado3}
I.\,A.~Ado, Private Communication (2015).

\bibitem{Alireza}
A.~Qaiumzadeh, R.\,A.~Duine, and M.~Titov, Phys.\ Rev.\ B \textbf{92}, 014402 (2015).

\bibitem{Tretiakov2008}
O.\,A.~Tretiakov, D. Clarke, G.-W. Chern, Y.\,B.~Bazaliy, and
O.~Tchernyshyov, Phys. Rev. Lett. \textbf{100}, 127204 (2008).

\bibitem{Clarke2008}%Dynamics of a vortex domain wall in a magnetic nanostrip: Application of the collective-coordinate approach
D.\,J.~Clarke, O.\,A.~Tretiakov, G.-W.~Chern, Ya.\,B.~Bazaliy, and O.~Tchernyshyov,
Phys.\ Rev. B \textbf{78}, 134412 (2008).

\bibitem{Tveten13}%Staggered Dynamics in Antiferromagnets by Collective Coordinates
E.\,G.~Tveten, A.~Qaiumzadeh, O.\,A.~Tretiakov, and A.~Brataas,
Phys.\ Rev.\ Lett. \textbf{110}, 127208 (2013).

\bibitem{Schuette14}%Inertia, diffusion, and dynamics of a driven Skyrmion
C.~Sch\"utte, J.~Iwasaki, A.~Rosch, and N.~Nagaosa, 
Phys.\ Rev.\ B \textbf{90}, 174434 (2014).

\bibitem{Barker2016}%Static and Dynamical Properties of Antiferromagnetic Skyrmions in the Presence of Applied Current and Temperature
J.~Barker and O.\,A.~Tretiakov, 
Phys.\ Rev.\ Lett. \textbf{116}, 147203 (2016).

\bibitem{Thiele}
A.\,A.~Thiele, Phys.\ Rev.\ Lett. \textbf{30}, 230 (1973).

\bibitem{Tretiakov2007}
O.\,A.~Tretiakov  and O.~Tchernyshyov, Phys. Rev. B \textbf{75}, 012408 (2007).

\bibitem{Ezawa2016} %High-topological-number magnetic skyrmions and topologically protected dissipative structure
X.~Zhang, Y.~Zhou, and M.~Ezawa, Phys.\ Rev.\ B \textbf{93}, 024415 (2016). 

\bibitem{Kadanoff}
L.~Kadanoff and G.~Baym, \textit{Quantum Statistical Mechanics}, W. A. Benjamin, Inc.: New York (1962).

\bibitem{Rammer}
J.~Rammer and H.~Smith, Rev.\ Mod.\ Phys. \textbf{58}, 323 (1986).

\end{thebibliography}
\end{document}